\documentstyle[11pt,aaspp4,epsfig]{article}

\lefthead{Herbst et al.}

\righthead{Rotation in the Orion Nebula Cluster} 

\begin{document} 
 
\title{Rotation in the Orion Nebula Cluster} 
 
\author{W. Herbst} 
\affil{Astronomy Department, Wesleyan University, Middletown, CT 06459}

\author{K. L. Rhode} 
\affil{Department of Astronomy, Yale University, New Haven, CT 06520} 

 \author{L. A. Hillenbrand} 
\affil{Astronomy Department, California Institute of Technology, Pasadena, CA 91125} 
\and
 \author{G. Curran} 
\affil{Astronomy Department, Wellesley College, Wellesley, MA 02181-8286} 
\begin{abstract} 

Eighteen small (4 arc-min square) fields within the Orion Nebula
 Cluster (ONC) have been photometrically monitored for one or more
 observing seasons between 1990 and 1999 with a CCD attached to the
 0.6 m telescope at Van Vleck Observatory on the campus of Wesleyan
 University. Data were obtained exclusively in the Cousins I band on
 between 25 and 40 nights per season. Results from the first three
 years of operation of this program were summarized and analyzed by
 Choi \& Herbst (1996). Here we provide an update based on six
 additional years of observation and the extensive optical and
 infrared study of the cluster by Hillenbrand (1997) and Hillenbrand
 et al.  (1998). Rotation periods with false alarm probabilities (FAP)
 $<$ 1\% are now available for 134 members of the ONC. Of these, 67
 were detected at multiple epochs with identical periods by us and an
 additional 15 were confirmed by Stassun et al. (1999) in their study
 of Ori OBIc/d.  Therefore, we have a sample of 82 stars with
 virtually certain rotation periods and another 52 with highly
 probable periods, all of which are cluster members. The bimodal
 period distribution for the ONC reported by CH is confirmed but we
 also find a clear dependence of rotation period on mass.  This
 phenomenon can be understood as an effect of deuterium burning, which
 temporarily slows the contraction and, therefore, spin-up of stars
 with M $\leq$ 0.25 M$_{\odot}$ and ages $\sim$ 1 My. Stars with M $<
 0.25$ M$_{\odot}$ have not had time to bridge the gap in the period
 distribution at around 4 days. Excess H-K and I-K emission, as well
 as CaII infrared triplet equivalent widths (Hillenbrand et al. 1998),
 show weak but significant correlations with rotation period among
 stars with M $>$ 0.25 M$_{\odot}$. Our results provide new
 observational support for the importance of disks in the early
 rotational evolution of low mass stars.

\end{abstract} 
 
 
\section{Introduction}

It has long been recognized that stars must lose an enormous amount of
angular momentum as they make the transition from a protostellar state
to the main sequence. When and how they do this is still uncertain
(cf. Bodenheimer 1995). The youngest visible stars, T Tauri stars,
are typically rotating at only about one-tenth of their critical
velocity, as first shown by Vogel \& Kuhi (1981) and later confirmed
by Bouvier et al. (1986) and Hartmann et al. (1986). Rotational
studies of the youngest nearby cluster - the Orion Nebula Cluster
(ONC) - indicate that about 2/3 of the low mass members are slow
rotators but the rest spin at higher rates, even approaching critical
velocity in the most extreme cases (Attridge \& Herbst, 1991,
hereinafter AH; Choi \& Herbst 1996, hereinafter CH; Stassun et
al. 1999, hereinafter SMMV). This wide range in spin rate and bimodal
character of the frequency distribution persists in clusters as old as
the Pleiades ($\sim 120$ My; Stauffer et al. 1994).

Many attempts have been made to model the rotational evolution of low
mass stars and account for the frequency distribution of rotation
periods in clusters of various ages (e.g. the recent work of Bouvier
1997, Krishnamurthi et al. 1997, and Barnes et al. 1999). Surface
angular momentum losses through magnetized winds and by redistribution
in the stellar interior are important processes which may act on
timescales of around 10$^7$ - 10$^8$ yrs. They can account for the
evolution of rotation from the T Tauri phase on. However, they cannot
explain the generally slow rotation and the existence of a wide range
of angular velocities in a cluster as young as the ONC. Much more
efficient angular momentum regulation mechanisms are needed which
operate on a timescale of $\sim$1 Myr or less.

The most commonly invoked angular momentum regulation mechanism during
the pre-main sequence phase is sometimes called ``disk-locking" and
based on an application by K\"onigl (1991) of the Ghosh and Lamb
(1979) theory of magnetic interaction between a rotating central star
and an accretion disk.  Somewhat similar proposals were made by
Cameron \& Campbell (1993) and Shu et al. (1994). In all cases, it is
possible to account for the typical rotation period of a T Tauri star
or slow rotator in the ONC ($\sim$8 days) with reasonable parameters
for magnetic field strengths, accretion rates, etc.  Edwards et
al. (1993) discussed evidence of a link between rotation rates and
disk indicators such as IR excess which they interpreted as support
for this disk-interaction model of angular momentum evolution. These
successes have led to a widespread adoption of some form of
disk-locking in most models of rotational evolution (cf. Li \& Cameron
1993; Cameron, Campbell \& Quaintrell 1995; Bouvier 1997; Krishnamurthi
et al. 1997; Barnes et al. 1999).

The ONC is a critical cluster for studying early angular momentum
evolution since it provides the largest sample of extremely young
stars in the solar vicinity. It has also been studied extensively by a
variety of techniques and masses, ages and infrared excesses
attributable to disk emission are available for about a thousand
cluster members (Hillenbrand, 1997; Hillenbrand et al. 1998). Rotation
periods for some members of the Trapezium Cluster, which is at the
heart of the ONC, were discovered by Mandel \& Herbst (1991). This was
followed by a more extensive study of the ONC by AH who found that the
period distribution in the cluster was bimodal. Further work, by CH,
confirmed that there are two peaks in the period distribution (near 2
days and 8 days) with a gap between them. CH interpreted this in terms
of the disk-interaction models. The long period peak is composed of
stars which are either currently interacting with their disks or have
just recently been released from them. The gap is caused by the rapid
spin-up expected for extremely young, contracting stars which are
evolving under conditions of angular momentum conservation. The peak
at 2 days is a binning phenomenon - there are relatively few bins
available for rapid rotators when period, rather than (say) angular
velocity is used as the independent variable. Recently, SMMV reported
rotation periods for 254 stars in Ori OB Ic/Id, including 104 stars
which are proper motion members of the ONC. They questioned the
statistical significance of the bimodal distribution and the mechanism
of disk-locking.

In this paper we summarize results on rotation in the ONC based on
nine years of monitoring at Van Vleck Observatory. The observations
are outlined in Section II, followed by a discussion of our period
detection techniques and results. With firmly established rotation
periods for 134 members of the ONC we are in a position to analyze, in
Section III, how rotation depends on mass, age, disk properties,
etc. This leads to a discussion, in Section IV, of what our data imply
for the rotational evolution of young, low mass stars. In contrast to
the results presented by SMMV, we find new support for a
disk-regulation model of angular momentum evolution and explain how
and why our results differ from theirs. Throughout the paper we rely
on the same pre-main sequence models adopted by Hillenbrand (1997) to
derive masses and ages, namely those of D'Antona and Mazzitelli (1994;
DM94).

\section{Observations, Reductions and Rotation Periods}

Data were obtained over 8 observing seasons from 1990/91 through
1998/99 with a CCD attached to the 0.6 m telescope of the Van Vleck
Observatory on the campus of Wesleyan University in Middletown,
CT. For the first 7 years, the camera employed a 512x512 chip; last
year it was replaced with a 1024x1024 chip. The descriptions and
statistics that follow refer to the first 7 years; only the field size
and locations have changed with the advent of the larger chip. Earlier
papers in this series have discussed the results in 1990/91 (Mandel \&
Herbst 1991), 1991/92 (AH) and 1992/93 (CH). A paper on the Trapezium
Cluster was also published by Eaton, Herbst \& Hillenbrand (1995). The
observing and reduction procedures are unchanged from what was
described by CH. Briefly, we obtain five consecutive one-minute
exposures in I from one to three times each clear night for each of
the fields in our program. The locations of the fields are shown in
Figure 1, which is centered on the Trapezium stars. Plus signs (+)
mark the location of the 1053 stars from Jones and Walker's (1988)
study that are within the area surveyed by Hillenbrand (1997). Filled
circles indicate stars with rotation periods discovered by us. At the
distance of the ONC (470 pc), 0.1$\deg$ $\sim 0.8$ pc, so the area
shown is roughly the extent of the dynamical ONC cluster modeled by
Hillenbrand and Hartmann (1998), $\sim$ 2 pc. The Trapezium cluster is
the inner $\sim 0.2$ pc, or about the size of the small central square
in Figure 1. According to Hillenbrand and Hartmann (1998) the
Trapezium cluster should be regarded as the core of the ONC rather
than as a separate entity. However, it does appear to be the youngest
portion of the cluster and is certainly the densest. Figure 2 gives
the naming scheme for our fields and Table 1 gives the Jones and
Walker (1988; JW) number for the central star of each field. The table
also indicates the years during which a particular field was
monitored, the number of stars for which we obtained photometry (N),
and the number of JW stars within the field. Although we monitored 69
stars in the fields which were too faint to be in the JW catalog, only
one of these was found to be periodic, probably because of increased
photometric errors (see below). Our most intensive monitoring has been
directed at the Trapezium cluster, which has been monitored every
year.

Aperture photometry is performed on all stars in the images using
standard IRAF tasks. Comparison stars are identified by an iterative
process which results in the selection of 4 or more non-variable or
slightly variable stars, whose average magnitude (not flux) defines
the comparison value. Typical errors are 0.015 mag for stars brighter
than about I = 14. We can only establish variability, therefore, for
stars with amplitudes exceeding about 0.05 mag. Although we find many
irregular variables we defer discussion of them to a later paper. Here
we discuss only those stars whose variations are strictly periodic in
at least one observing season.

The periodogram technique discussed by Scargle (1982), as implemented
by Horne \& Baliunas (1986), was used to search for periodic
variables. All stars were examined independently in each season, and a
combined periodogram was formed by multiplying those obtained in
individual seasons. The combining procedure did not reveal any stars
which we would have missed by looking only at individual seasons, but
it did yield some potentially interesting information about a few
stars which is described in the footnotes to Table 2.  The false alarm
probability (FAP) defined by Horne \& Baliunas (1986) was used to
identify stars with likely periodic variations. A threshold value of
1\% was employed when only a single observation per night was
available or when multiple observations per night were averaged. As
discussed by Herbst \& Wittenmyer (1996), the standard FAP
prescription assumes uncorrelated data and that may not be a valid
assumption when there are multiple observations per night. In fact,
one will obtain highly misleading FAP estimates if the standard
prescription is used under these conditions, which will lead to the
false identification of rotation periods for stars whose variations
are probably random. To avoid this problem, we use a Monte Carlo
method (cf. Herbst \& Wittenmyer, 1996) to get a more realistic
estimate of the FAP. In any event, our final test is to examine the
light curves for all stars with possibly real periods. If the light
curve is not convincing, we reject the star as a real periodic
variable despite what the FAP statistic may indicate. In practice,
only a few stars are rejected because of their light curves and we
find the Horne \& Baliunas prescription, as well as our Monte Carlo
method, to be fairly reliable.

Even when a reliable statistic is used, however, it must be expected
that about 1\% of the stars searched for periodicity will, in fact,
turn out to be false alarms. The advantage of repeating this work in
multiple seasons is that we can gradually isolate a sub-sample which
is free of false alarms by finding stars which are periodic in more
than one season. The chance of a false alarm occurring twice for the
same star (with the same period!) in different seasons is, of course,
negligible. Table 2 summarizes our results after 8 seasons of
observation. The stars are divided into two groups in what follows -
those found to be periodic at more than one epoch (including those
with matching periods reported by SMMV) and those found to be periodic
in only a single season. The first group can be regarded as consisting
of stars whose periods are established with certainty while the second
group could still contain false alarms.  It is gratifying to note that
in virtually all cases, when a star was found to be periodic in two
different years, the periods were identical to within the errors. The
only exceptions were a couple of cases in which an harmonic or an
alias of the period was found (see notes to Table 2). In all, there
are now 82 stars with firmly established periods (multiple season
detections) and an additional 52 with likely periods (single season
detections).

Light curves for some of these stars have already been displayed by CH
or AH. We show light curves for the rest in Figure 3. In all cases,
the periods, amplitudes and shapes of the light curves indicate that
these are spotted variables whose periodicity is caused by their
rotation. We identify the period as the rotation period of the star. A
completely independent check of our results is now provided by the
work of SMMV.  We have 47 stars in Table 2 which are in common with
them. In 44 cases the periods reported by SMMV agree with ours to
within the errors of the determination. This is a gratifying
comparison for both studies and reinforces our confidence in the
methodology and basic result that we are, indeed, measuring rotation
periods for these stars. Since the SMMV time span is rather short (at
most 17 days and, in many cases, only 11 days) their periods are more
uncertain than ours, so we do not include their values in our
averages. We do, however, consult their data when the stars are
divided into single and multi-epoch detections.

The cases in which significantly different periods are reported are as
follows (with our period and the SMMV period given, respectively): JW
379 (5.62, 11.3), JW 843 (5.38, 0.84) and JW 984 (8.44, 1.13). JW 379
is a case in which our period is one-half of that reported by
SMMV. They suggest that this may be an example of period doubling
caused by spots on opposite hemispheres of the star. Alternatively,
the star may have changed its spot characteristics during the single
cycle over which they observed it. Their light curve clearly shows two
minima. Here we adopt the longer period, recognizing that additional
photometry is needed to be certain about which period is the proper
estimate of the star's rotation period. We, therefore, do not regard
the star as a multi-epoch detection in what follows.  The two periods
(P) found for both JW 843 and 984 are examples of ``alias pairs"
discussed by CH. They are related by 1 $\pm$ 1/P, i.e. the beat
frequency with a one day sampling interval. Since SMMV have better
time resolution and have data obtained at widely separated longitudes
they should be less sensitive to aliasing problems. However, the light
curve shown by SMMV for the 1.13 day period of JW 984 is not
convincing, especially given the short time interval over which they
observed the star. We, therefore, have adopted the longer period for
this star but, to emphasize the uncertainty have not included it among
the multi-epoch period detection sample. Obviously, further data is
necessary to resolve its period. The same is true of JW 843, which was
only a single epoch determination to begin with and, in light of the
uncertainty over its true period, we continue to regard as such.

\section{Results}

\subsection{The Periodic Sample}

Of the 134 periods discovered as a result of the VVO monitoring
program, 133 of them have JW proper motions, and only one star (JW
794) has a membership probability less than 50\%. In most cases, the
proper motions indicate membership in the ONC for our periodic stars
at the 99\% confidence level.  Figure 4 shows the color-magnitude
diagram for all of the JW stars monitored in our fields, with the
periodic stars marked by solid circles and the rest by plus signs. It
is apparent from this diagram that our limit for detecting periodic
stars is about I $\sim$ 16, although we monitored fainter objects. It
is also apparent that most of our periodic stars are brighter than I
$\sim$ 14, and below that we presumably have an increasingly difficult
time detecting periods. This is almost certainly due to the fact that
the errors of the photometry, especially for stars embedded in
nebulosity, become larger at those magnitudes making it more difficult
to find periods, especially for stars with smaller amplitudes. We note
that, in addition to the 665 JW stars in our fields, we monitored 69
stars which were not included in the JW survey because of faintness,
but are included in Hillenbrand's (1997) study. Only one of those
stars, number 3128, was found to be periodic. Essentially then, our
period monitoring program can be considered limited to the JW sample
and somewhat incomplete below I $\sim$ 14 mag. For stars brighter than
I = 14 mag, we detected periods in 33\% of those monitored. Except for
the OB stars, none of the objects in our fields are saturated, so
there is no incompleteness on the bright end. There is an apparent
paucity of periodic stars with I mag brighter than 10, which may be
real.  Altogether, we have monitored about 63\% of the JW stars in the
ONC; this number is fairly independent of magnitude range. It is clear
from Figure 1 that our concentration is towards the center of the
cluster, where the fraction of monitored stars which are cluster
members is presumably highest.

Every periodic star in our JW sample has had its mass derived by
 Hillenbrand (1997) and, in Figure 5, we show their distribution
 (bottom panel) compared to the mass distribution of all JW stars
 within the monitored fields (top panel). It is clear that
 uncertainties in the photometry for fainter stars limit our ability
 to find periods for the lowest mass stars (M $<$ 0.25
 M$_{\odot}$). The mass distribution of periodic stars peaks between
 0.25 and 0.5 M$_{\odot}$ whereas for the cluster as a whole, the mass
 distribution continues to rise well below our limit for detecting
 periods. In the periphery of the ONC it should be possible to push
 the period detections to fainter, lower mass stars by using larger
 telescopes or longer exposure times. However, in the center of the
 cluster this may prove difficult (at least from the ground) because
 of the nebular background brightness.

\subsection{The Bimodal Period Distribution and its Dependence on Mass} 

The frequency distribution of rotation periods in the ONC was
discovered to be bimodal by AH and the result was confirmed by CH, but
recently questioned by SMMV. In Figure 6 we show the period
distribution based on the current sample, which is 75\% larger than
that analyzed by CH. The bimodal nature described in the earlier work
remains clearly visible, with peaks near 2 days and 8 days and a gap
at 4 days. The small tail of more slowly rotating stars out to at
least 20 days and possibly 34 days is also apparent. We test the
statistical significance of the gap below, but first discuss the
dependence of rotation on mass.

Figure 7 shows the distribution of mass with rotation period for our
sample. Multi-epoch detections are shown as solid circles and
single-epoch detections as crosses. An enlarged view of the boxed area
in the top panel is shown in the bottom panel. It is immediately
apparent from this plot that the lower mass stars (M $<$ 0.25
M$_{\odot}$) have a more uniform distribution of periods than do those
of higher mass. In particular, the gap in the period distribution
around 4 days, which is quite obvious for stars in the range 0.25 $<$
M $<$ 1.0, is closed for stars less massive than 0.25
M$_{\odot}$. This may be seen more clearly in Figure 8, where we show
histograms of the period distribution for the higher and lower mass
stars respectively. Two principle differences may be noted. First, the
gap at 4 days is obviously quite strong in the higher mass
distribution but absent among the lower mass stars.  A more subtle,
but nonetheless apparent difference in the distributions is the
relative lack of stars with very short periods among the lower mass
stars. As is shown below, both of these features can be understood in
terms of mass-dependent rotational evolution.

Although the period distributions shown by CH and in Figures 6 and 8
of this paper, are obviously bimodal, the statistical significance of
the gap at 4 days was recently questioned by SMMV. They claim that the
CH distribution between 1 and 10 days is not significantly different
from a uniform distribution. The basis for this is a
Kolmogorov-Smirinov (K-S) test which shows essentially no difference
between a ``model" uniform distribution and the actual distribution at
about the one sigma level. The problem with this is that the K-S test
is not sensitive to gaps in the data, as SMMV themselves
note. Obviously, therefore, the K-S test, and other cumulative
distribution analyses like it, are inappropriate for evaluating the
reality of gaps. They will give misleading results if applied and
believed.

Fortunately, a simple, appropriate statistical test for bimodality
does exist - the Double-Root Residual (DRR) method discussed in an
astronomical context by Gebhardt \& Beers (1991) and Ashman, Bird \&
Zepf (1994). This involves calculation of a quantity, the DRR, defined
as:

$$DRR=\sqrt{2 + 4 N} - \sqrt{1+ 4 A}$$ 

for all non-zero values of N, where N is the number of objects in a
particular bin and A is the average number of objects in a bin. The
advantage of this statistic is that it indicates exactly where the
data and the model differ from each other and gives directly the
significance level of the difference. That is, a value of DRR=2, for
example, corresponds to a discrepancy between the data and the model
at the 2$\sigma$ (95\%) confidence level. Technically, the DRR test
does not test for bimodality {\it per se} but for the significance
level of features such as peaks and a gap. If a single significant gap
is found between two relative peaks, however, the logical inference is
that the distribution is bimodal.

In Figure 9 we show three truncated period distributions which are
tested for significant gaps using the DRR method. We chose the
interval 1 - 9 days as a convenient compromise between the 1-10 and
1-8 day uniform distribution ``models" which SMMV test. The top panel
shows the CH distribution as modified by SMMV, which they claim is not
significantly different from a uniform distribution based on a K-S
test. The DRR test, in the top right panel, indicates otherwise. The
gap around 4 days is significant at about the 3$\sigma$ (99\%) level,
by itself, and when the positive peaks on either side are considered,
the statistical significance of the gap is enhanced. The middle and
lower histograms are for the full sample of periods discussed in this
paper and for the sample with M $>$ 0.25M$\odot$. In all cases, the
DRR test indicates that the gaps are real and that the distributions,
therefore, may be regarded as bimodal, both qualitatively and
quantitatively.

\subsection {Correlations of Rotation Period with Position in the Cluster, IR excess and CaII emission}

We have searched for correlations between the rotation period and
other measurable characteristics of the stars, and have found four. In
all cases, the correlations are relatively weak, but Spearman
Rank-Order Correlation Tests (e.g. Press et al. 1986) indicate they
are significant at the $\sim$ 99\% confidence level or better. The
data, which are taken from Hillenbrand (1997, 1999) and Hillenbrand et
al. (1998) are listed in Table 3, the four correlations are displayed
in Figures 10 to 13, and results of the statistical tests are given in
Table 4. These include the Spearman correlation coefficient, the
probability that this correlation would arise by chance in an
uncorrelated data set, the slope of a linear least squares fit to the
data and the 1$\sigma$ error of the slope. Since there is no reason to
expect a linear relation between rotation period and any of the other
quantities, the non-parametric Spearman statistic should be the more
reliable indicator of a correlation. It also has the virtue of
depending only on the rank order of the data, not on actual values, so
it is a more robust test than the least squares fits in the sense that
one or two extreme values (such as the unconfirmed period at 34 days)
will have little effect on the results. It may be seen from Table 4
that the Spearman tests indicate uniformly higher significance levels
than do the linear least squares fits, but even these indicate
significant slopes for all the correlations except H-K at the $\sim$
95\% confidence level or better. Since rotation periods are derived in
a manner which is {\it totally independent} of the other parameters,
and since there are many possible ways in which correlations such as
these could be destroyed, but none (which we can think of) by which
they could be falsely created (i.e. other than by a real physical
process), we find them impressive in spite of their relative
weakness. Added to this is the fact that all four correlations have
simple physical interpretations which may account both for their
existence and their sense.

Figure 10 shows the correlation that exists between rotation period
and location within the ONC, measured by the projected radius (in
arc-minutes) from the cluster center ($\Theta^1$ Ori; Hillenbrand
1997). As Table 4 indicates, the correlation is quite significant,
even though the correlation coefficient is fairly small. This remains
true even if we exclude the star with a rotation period of 34 days. It
is also apparent that the correlation exists among the sample of
multi-epoch stars by themselves. A similar correlation was noted by CH
and by Eaton, Herbst \& Hillenbrand (1995), who showed that the period
distribution in the Trapezium cluster (the central $\sim$ 4
arc-minutes of the ONC) included more stars with longer periods than
was the case in the remainder of the ONC. While the Trapezium cluster
has sometimes been regarded as a separate entity (e.g. Herbig and
Terndrup, 1986), Hillenbrand and Hartmann (1998) consider it to be
merely the center of the ONC. However, Hillenbrand (1997, 1998) has
shown that there is an age gradient within the ONC in the sense that
the center is younger, and that the fraction of stars with disks, as
indicated by IR excess emission, is higher towards the center. The
correlation presented here, therefore, associates longer rotation
periods with the portion of the ONC which is both younger and contains
more stars with disks. Since contracting stars will spin faster as
they age, in the absence of an angular momentum loss mechanism, the
correlation with position in the cluster may simply reflect the
extreme youthfulness of the center of the ONC relative to its outer
parts. Alternatively (or, in addition) if the disk-interaction
mechanism is operating to drain angular momentum, then stars with
disks may, in general, rotate more slowly than those without
disks. Either, or both, physical effects could account for the
existence and sense of the observed correlation.

Figures 11 and 12 show two measures of IR excess and their correlation
with rotation period. The I-K excess is taken from Hillenbrand et
al. (1998) and has the advantages of being available for more stars
and involving a longer color baseline, which should make it easier to
identify significant disk emission. However, the ubiquitous
variability of the stars in I and the non-simultaneity of the measures
at different wavelengths increases the scatter in the
data. Nonetheless, a significant correlation between rotation period
and I-K excess does exist among stars more massive than 0.25
M$_{\odot}$ in the sense that stars with larger IR excesses tend to
have longer rotation periods. Again, removing one or two extreme stars
has little effect on the Spearman test results (although it can change
the least squares slopes more dramatically). The same results are
obtained when the H-K excess (Table 3) is examined. In this case, the
scatter is smaller, presumably because of the lower variability of the
stars in H and the near simultaneity of the measurements at different
wavelengths. On the other hand, there are fewer stars for which these
data are available and the color baseline is smaller. The net result
is that a correlation exists with about the same (Spearman)
coefficient and at about the same significance level as for I-K,
although the least squares solutions do not indicate a significant
slope. Taken together, these figures indicate a reasonable likelihood
that rotation period is, in fact, correlated with the presence of disk
emission. This supports the results of Edwards et al. (1993) who
presented a similar correlation among a group of stars which partially
overlaps with our sample, but is mostly independent of it. The
physical interpretation proposed by Edwards et al. (1993) was that
disks (can) act to slow the rotation of stars through magnetic
interaction (K\"onigl 1991; Ostriker \& Shu 1995); hence, stars with
disks may either be still ``locked" to them or just recently released
and will, in either case, tend to be slower rotators than stars whose
disks were dissipated at earlier times.

Finally, in Figure 13, we show one additional correlation which exists
and is important in spite of its weakness. This is between rotation
period and equivalent width (EW) of the CaII infrared triplet
(negative values indicate emission) as measured by Hillenbrand et
al. (1998). This is an entirely independent measure of disk
``strength", presumably reflecting the current accretion rate. Again,
the sense of the correlation among the stars more massive than 0.25
M$_{\odot}$ is consistent with what one would naively expect if the
disk-interaction scenario of Edwards et al. (1993) were valid. Stars
with rotation periods shortward of the gap in Figure 13 almost all
have CaII in absorption, indicating weak or non-existent
accretion. Stars with rotation periods longward of the gap may also be
in absorption, but have an average EW (CaII)$=-0.03 \pm 0.38$, which
is significantly less than the corresponding value ($1.49 \pm 0.32$)
for stars shortward of the gap. As was the case for IR excess
emission, it is difficult to conceive of observational circumstances
or selection effects which could conspire to create a correlation such
as this in our data set, by chance, while it is easy to imagine how
the correlation might be destroyed, even if the disk-locking mechanism
were operative.  In addition to random errors and false alarms among
the periods, there could be complexities in the disk-interaction
process which might obviate naive expectations about even the sense of
the correlations. For example, higher accretion rates are expected to
cause the radius of co-rotation to shrink and the rotation period,
therefore to decrease in disk-locking models. This would produce a
trend in the data which is counter to that expected in a simple
``locked" or ``not locked" scenario. The existence and sense of the
observed correlations are, therefore, more interesting to us than the
relatively small values of the correlation coefficients. We turn now
to a discussion of the results in terms of models of rotational
evolution.
 
\section {Discussion}

\subsection {Understanding the Dependence of Rotation on Mass}

Perhaps the most interesting new result described above is the
dependence of rotation period on mass, illustrated in Figures 7 and
8. Can we understand why the gap in the period distribution disappears
for stars less massive than about 0.25 M$_{\odot}$ and why there are
substantially fewer very rapid rotators among these lower mass stars?
To address this issue we consider the expected behavior of rotation
period with mass and age under conditions of angular momentum
conservation, based on the models of DM94. For definiteness we adopt
an initial rotation period of 10 days for a star at an age of 0.07
My. Changing the initial period by some scale factor is equivalent to
adjusting the entire period evolution by the same factor - the shapes
of the curves remain the same. The evolution of period with age for
four different masses is shown in Figure 14. We have assumed
homologous contraction for these entirely convective stars, so the
period (P) depends only on the radius (R) as P $\propto$ R$^{2}$. The
rapid decline in period during the first few hundred thousand years
reflects the rapid contraction of these stars during the earliest
phases of PMS evolution.

At first, there is no difference between the evolution of higher and
lower mass stars. However, by an age of $\sim$ 1 Myr (typical of the
ONC) the lower mass stars are rotating with much longer periods than
the higher mass stars. The physical basis for this is the so-called
deuterium burning ``main sequence". Low mass PMS stars are expected to
survive their protostellar evolution with most or all of their initial
deuterium abundance intact (Palla and Stahler 1999). When the
interior temperatures get high enough, deuterium burning is initiated
and the contraction towards the main sequence is temporarily
slowed. Since P is quite sensitive to R, this evolutionary effect
becomes quite obvious in the rotation period tracks of the lower mass
stars. Stars more massive than $\sim$ 0.3 M$_{\odot}$ go through
deuterium burning too rapidly and at too early a time for it to be
important to their observable period evolution. This may even happen
during the protostellar stage (Palla \& Stahler 1999) rather than the
PMS stage. After $\sim$ 2 Myr the difference between the masses has
largely been erased.

We find it intriguing that this simple, heuristic model can account
for the observed difference in period distributions between the higher
and lower mass stars in the ONC. Stars with M $<$ 0.25 M$_{\odot}$
which have evolved without disk locking should have periods of around
4 days, whereas more massive stars should have periods of around 2
days. In other words, the lower mass stars are expected to lie within
the ``gap" defined by the period distribution of the higher mass stars
- precisely as we observe. If there is an age range in the cluster, as
expected, the oldest low mass stars might have had time to spin up to
periods near two days, but the most rapidly rotating lower mass stars
will have significantly longer periods than the higher mass stars
until ages of about 2-3 Myr. We can, therefore, also understand on the
basis of this scenario the relative lack of very short periods among
the low mass stars, assuming the oldest of them is about 2-3 Myr. Note
that starting radii for our calculations are about equal to the
birthline radii for stars of the same mass (e.g. Stahler 1999) so the
rapid period evolution expected during the early PMS phase is not
obviated by the consideration of a finite starting time or radius
corresponding to the protostellar stage.

\subsection{The Disk Locking Hypothesis}

AH first suggested and CH established in more quantitative fashion
that the disk locking hypothesis of K\"onigl (1991) and/or Ostriker
and Shu (1995) provided a theoretical framework within which one could
understand the rotation period distribution in the ONC and, in
particular, its bimodal nature. Longer period stars are interpreted as
being still locked or only recently released from their disks. Shorter
period stars are interpreted as having been released early in their
PMS evolution to spin up in the fashion depicted in Figure 14. The end
of the disk-locking phase could result from a cessation of accretion
or from a weakening of the surface magnetic field, if it were
primordial, or other factors. The correlation between IR excess
emission and rotation period shown by Edwards et al. (1993)
strengthened the argument for disk regulation. The enlarged data set
and analysis presented here provides further support for this
scenario, as we now discuss.

The strongest argument in favor of disk locking is the bimodal period
distribution of stars more massive than 0.25 M$_{\odot}$ (Figure
8). This is now established both qualitatively and quantitatively. The
gap is a result of the rapid period evolution of young PMS stars which
are conserving angular momentum. Figure 14 clearly shows how rapidly a
star with M $>$ 0.25 M$_{\odot}$ will reach a period near 2 days under
those circumstances. The peak at 2 days is populated with stars which
have contracted under angular momentum conservation from somewhere
near their starting points in the DM94 tracks or, equivalently, from
the birthline (Stahler, 1999). The existence of a peak near 8 days is
difficult to understand without recourse to the disk-locking
mechanism. As CH pointed out, models such as those of Ostriker \& Shu
(1995) naturally yield a disk-locking period of about 8 days for
reasonable values of the parameters. As far as we are aware, no other
theory of rotational evolution can account for the period distribution
either quantitatively or qualitatively. Since stars of all masses
would be expected to pass through the 8 day period regime in very
quick order (cf. Fig. 14), the existence of the peak in the frequency
distribution at this period indicates that some sort of period
regulation mechanism must be operating.

Johns-Krull, Valenti \& Koresko (1999) have recently calculated the
average stellar magnetic field, B (at the magnetic equator), required
by disk-locking models given observed masses, radii, mass accretion
rates and rotation periods for T Tauri stars. Following them (and CH)
we calculated B for the ONC stars, using masses and radii from
Hillenbrand (1997), rotation periods from this study, and assuming a
uniform mass accretion rate ($\dot M$) of $10^{-8}$ M$_{\odot}$
yr$^{-1}$. We adopt the model of Ostriker \& Shu (1995) but, as
Johns-Krull, Valenti \& Koresko (1999) show, the models of K\"onigl
(1991) and Cameron \& Campbell (1993) give similar results to within a
factor of two. In our case, $$B = 1068 {{M^{5/6}P^{7/6}} \over {R^3}}
$$ where B is in Gauss, M and R in solar units and P in days. The
results are plotted in Figure 15 as a function of rotation period. For
clarity, the figure was cropped such that a few stars with very large
values of B (up to 13 kG) are not shown. They are stars with
particularly small radii. The important point, however, is that a
typical star in the ONC, with mass $\sim$ 0.3 M$_{\odot}$, radius
$\sim$ 2 R$_{\odot}$ and P $\sim$ 8 days, requires B $\sim$ 500 Gauss
for disk-locking if the mass accretion rate is what we assumed. Since
B $\propto \sqrt{\dot M}$ this result does not change much if the
accretion rate is somewhat in error.

Values of B $\sim$ 500 G for pre-main sequence stars are quite
reasonable to assume, even though direct evidence for them is very
difficult to obtain. The classical T Tauri star, BP Tau, has a field
of $2.6 \pm 0.3$ kG according to Johns-Krull, Valenti and Koresko
(1999). The existence of spots on the ONC stars implies locally strong
fields capable of either channeling accreting matter onto hot spots or
disrupting convection to cause cool spots. On the Sun, the field in
dark spots is $\sim$ 1-4 kG. If the field on the ONC stars is also
concentrated into (two polar) dark spots, we would expect similar
field strengths, since the filling factor for the spots is $\sim$ 10 -
30\% based on the photometric amplitudes. We conclude that the
disk-locking models are quantitatively reasonable in their magnetic
field and accretion rate requirements.

\subsection{Differences Between Our Study and That Of SMMV}

SMMV recently studied the rotation period distribution in and around
the ONC and obtained results which are in marked contrast to those
reported here. In particular, they found no evidence for a gap in the
period distribution and no correlation between rotation period and IR
excess emission. This led them to question the observational basis for
the disk locking hypothesis. They concluded, in conservative style,
that they did not find evidence that disk locking is the dominant
mechanism in angular momentum evolution during the PMS phase. In this
section we discuss the differences between our work and theirs which,
we believe, lead to different results and conclusions.

There are a variety of selection effects which result in differences
between our sample of stars with rotation periods and that of
SMMV. First, their study includes a much larger area around the ONC
than ours. In fact, it extends well beyond the boundaries of the ONC
as defined dynamically by Hillenbrand and Hartmann (1998) into the
region known as Ori Ic, an older subgroup of Ori OBI (Warren \& Hesser
1977). Second, they are missing the very center of the ONC - the
region of the Trapezium cluster - because of problems with
nebulosity. Third, their images are a bit deeper than ours, leading to
better signal to noise for the fainter stars at the expense of
saturation of the brighter ones. This results in their having probed a
different mass range than is the case for us. For example, only 46\%
of the stars in their sample for which they give masses have M $>$
0.25 M$_{\odot}$, while 61\% of the stars in our sample fall into that
mass range. They also have no stars with M $\geq$ 0.9 M$_{\odot}$
whereas 17\% of our sample is more massive than that. Finally, while
their total sample of stars in Ori OBIc and OBId is 254 stars, the
portion of the sample which is inside the defined boundaries of the
ONC is only 104 stars, and the number of those with M $>$ 0.25
M$_{\odot}$ is only 52 stars. Therefore, the fraction of their sample
which should show bimodality according to our results could be as low
as 20\% (i.e. 52/254), whereas in our sample it is 61\%. This may be
why bimodality is readily apparent in our data set, but less so in
theirs.

In addition, there are differences in the observational material and
its analysis which affect the samples. Their study was conducted over
a period of, at most, 17 days and, in some regions less. They believe
that incompleteness in their periods begins at about 8 days. Our
observations span $\sim$ 150 days in each of the 8 epochs so we have
no incompleteness at long periods. We also have multiple observations
per night, so we are sensitive to short periods as well. However, SMMV
do have the advantage of observations at different longitudes, making
them less susceptible to the beat phenomenon associated with a one day
observing interval. A very important difference, we believe, is that
82 of our 134 periods are ``multi-epoch", which helps us define a
subset of the data in which the chance of there being ``false alarms"
is miniscule. The percentage of the SMMV sample which is actually
false alarms is expected to be about 10\%, assuming that their method
of estimating FAP is accurate. (That is, in a sample of 2279 stars
which they searched for periods, about 23 would produce light curves,
by chance, which had FAP $<$ 0.01. This represents about 10\% of the
$\sim$ 230 stars with actual periods.) In fact, if their FAP
calculation is optimistic, the false alarm pollution could be
higher. With only single epoch data there is no way to check these
results at the present time, but an observational program to extend
the Wesleyan studies to a larger area around the ONC is underway.

Given the selection effects, it is not hard to account for the
different conclusions regarding the actual bimodality of the ONC
period distribution. Based on this study, we would only expect to find
a bimodal distribution in that subset of the SMMV data which has M $>$
0.25 M$_{\odot}$ and is as young as a typical ONC member, which could
be only 20\% of their sample, as mentioned above. Do we, in fact, find
a bimodal distribution in the portion of the SMMV data set where it is
expected according to our results? In Figure 16 we show the period
distribution for the 52 stars in the SMMV sample that are within the
boundaries of the ONC described by Hillenbrand (1997) and are more
massive than 0.25 M$_{\odot}$. The distribution is obviously bimodal
and entirely consistent with the results presented earlier in this
paper. The solid portion of the histogram indicates those stars from
the SMMV sample which have confirmed periods from this study. Clearly
there is no conflict in results. Both our data and the SMMV data
indicate a bimodal distribution for ONC members with M $>$ 0.25
M$_{\odot}$.

The fact that our entire sample continues to show bimodality, as did
the samples analyzed by AH and CH, while the SMMV sample does not (at
least as clearly), is a consequence of the different selection effects
mentioned above. In particular, their full sample of 254 stars
definitely has a greater proportion of low mass stars than ours and
probably has many more older stars from the Ori Ic subgroup. They are
also missing the youngest portion of the ONC (the Trapezium cluster)
and are incomplete for stars with periods exceeding 8 days. Finally,
they have a false alarm contamination of at least $\sim$10\%. All of
these factors tend to diminish the appearance of bimodality in the
SMMV total sample relative to the VVO sample.
 
The fact that SMMV failed to find correlations of period with IR data
(or CaII) is a result of the same selection effects discussed above
plus lack of a large enough sample size. Again, when we compare
similarly selected sub-samples we find no disagreement between the
results indicated by the two studies. In particular, for SMMV stars
with M $>$ 0.25 M$_{\odot}$ Spearman rank-order tests indicate
correlation coefficients that are about the same as what were found by
us and the least squares fits indicate identical slopes to within the
errors. The main difference is that the significance of the
correlations and slopes is much less for the SMMV data on account of
their smaller sample size (about 46 stars compared to our 69) and the
fact that they have a smaller range in period (less than 12
days). Including the $\sim$20 stars in the SMMV sample with IR excess
and/or CaII data which were not already in our sample has very little
effect on the correlations or significances shown in Table 4. We
conclude, once again, that there is no conflict between our data sets.

We take this opportunity to comment on two other aspects of the SMMV
paper which relate to the disk-locking hypothesis. First, in their
Figure 15, SMMV show a correlation of period distribution with
``age". Age is placed in quotation marks because it is highly model
dependent and, as Hillenbrand (1997) has noted, likely to be
unreliable for values less than 1 Myr owing to the neglect of the
birthline in the DM94 models (cf. Palla \& Stahler 1999). In fact, the
ages derived by Hillenbrand (1997) are mass dependent, as she
discussed and we show in Figure 17 for the mass range relevant to the
SMMV sample. Rather than sorting the stars by age, we believe that
SMMV have actually (inadvertently) sorted the stars by mass. Their
sample labeled log age $<$ 5 consists almost exclusively of stars with
M $\leq$ $\sim$0.3 M$_{\odot}$, whereas their sample labeled log age
$>$ 5.7 contains stars with the full range of masses. Our
interpretation of their Figure 15 is that it says nothing about the
evolution of period with age, since the age is not accurately measured
by the plotted quantity. But it does reflect the same trend of period
with mass displayed in our Figure 8, namely, the higher ``age" (mass)
sample appears bimodal while the lower ``age" sample is more uniform.

Our second comment on SMMV results refers to their discovery of stars
rotating at or near critical (breakup) velocity in the ONC. Although
these short periods and the membership of the stars must be confirmed,
it is not clear to us why SMMV regard them as posing a serious
challenge to the present picture of rotational evolution. The evidence
for disk locking discussed here relates to an event which happens in
the first $\sim$ 1 Myr of the star's evolution. As Figure 14
demonstrates, this is the critical time to impact the spin-up of the
star, since this is when the radius is changing most rapidly. If some
stars emerge from their protostellar stage spinning at critical
velocity and are not subsequently slowed by interaction with a disk,
then they may end up on the main sequence spinning at nearly critical
velocity as well. Of course, the well known mechanism of angular
momentum loss through magnetized stellar winds will act, on the longer
timescales of PMS contraction, to drain some angular momentum.
Therefore, as the star contracts, it will remain at, or near, critical
velocity but this poses no problem to a picture of rotational
evolution which involves disk-locking in the first few Myr for many
(but, of course, not all) stars. In fact, if all stars were
disk-locked we would be unable to understand the wide variation in
rotation periods observed in the ONC or in the Pleiades. So the
existence of stars with very short periods is in no way a challenge to
the standard scenario of rotational evolution which involves a brief,
but important, disk-locking phase as well as wind losses on a longer
time scale. Of course, the evolution of a star which is rotating at
critical velocity and still contracting is an interesting theoretical
issue. Perhaps such a star will be surrounded by an extruded disk
analogous to those seen in Be stars and perhaps these could, on
occasion be mistaken for accretion disks.

\section{Summary and Conclusions}

Eight years of monitoring stars in the ONC has led to the
determination of 134 rotation periods, 82 of which are confirmed in
multiple seasons. We find that rotational properties depend on
mass. Stars more massive than 0.25 M$_{\odot}$ have a bimodal
distribution with a significant gap near 4 days, while less massive
stars have a narrower distribution without a gap. The difference can
be understood in terms of stellar evolution. Lower mass stars with
ages $\sim$ 1 My are undergoing significant deuterium burning which
keeps them more inflated and, hence, spinning slower than the more
massive stars. The bimodal distribution for the more massive stars,
which was discovered by AH and confirmed by CH, but questioned by
SMMV, is established here both qualitatively and
quantitatively. Differences between our results and those of SMMV can
be understood in terms of selection effects.

Our data provide new observational support for the importance of
disk-locking in the angular momentum evolution of stars during their
first 1-2 million years. The strongest evidence is the bimodal
distribution itself. In the absence of a controlling factor such as
disk-locking, PMS stars more massive than 0.25 M$_{\odot}$ would
rapidly spin up to periods of 2 days or shorter, unless their initial
rotation periods from the proto-star phase were incredibly long
($\geq$ 40 days). The magnetic locking models of K\"onigl (1991) and
Ostriker \& Shu (1995) and evolutionary scenarios of Cameron,
Campbell, \& Quaintrell (1995) account for the 8 day peak in the
period distribution with reasonable parameters (e.g. surface magnetic
field strengths, accretion rates, etc.) which are on fairly solid
observational and theoretical grounds (cf. Johns-Krull, Valenti \&
Koresko 1999). It is the only theory that makes quantitative
predictions of this sort which can be checked by our
data. Disk-locking is observed to occur in analagous astrophysical
situations (e.g. dwarf novae; c.f. Sion 1999) and it is certainly the
leading candidate at present to explain the rotation period data in
the ONC and other young clusters.

Some additional evidence in support of disk-locking comes from four
weak, but significant, correlations which we find. Rotation period
correlates with position in the cluster in the sense that longer
periods are concentrated towards the central (``Trapezium cluster")
region. Hillenbrand (1997, 1998) showed that this is the youngest part
of the ONC and also contains the greatest percentage of stars with
disks. Among the stars more massive than 0.25 M$_{\odot}$, we also
find that rotation period is correlated with IR excess emission in
both I-K and H-K and is anti-correlated with CaII equivalent
width. All three findings indicate an association between rotation and
accretion disks similar to what was found by Edwards et al. (1993). As
discussed by those authors, this evidence supports the idea of
disk-regulated rotation.

\paragraph{Acknowledgments.} 

It is a pleasure to thank the W. M. Keck Foundation for their support
of the Keck Northeast Astronomy Consortium, which brought one of us
(G.C.) to Wesleyan as a summer student to participate in this
research. We thank J. Stauffer, R. Mathieu and K. Stassun for helpful
discussions of this work. It is also a pleasure to thank the many
Wesleyan students who participated in this work as observers, under
the able direction of Nancy Eaton, Lisa Frattare and Eric Williams. We
particularly thank Jamison Maley, Hugh Crowl, Frank Muscara, Lael
Hebert, Emily Lu, Arianne Donar, Eli Beckerman, Karen Kinemuchi,
Kristin Kearns, Janice Lee, Adam Heinlein, Aaron Steinhauer, Anil
Seth, Ben Holder, Stuart Norton, Chris Mazzurco, Kristin Burgess,
Rafael Verdejo, Phil Choi, Doug McElroy, Robin LeWinter, Anastasia
Alexov, Greg Keenan, Greg Vinton, Greg Mandel, Jamie Treworgy, Chuck
Ford, Elan Grossman, Andrew Billeb and Jody Attridge. NASA has
supported this research through grants from its ``Origins of Solar
Systems" program to two of us (W.H. and L.A.H.).
 
\subsubsection*{References.}
\noindent Ashman, K. A., Bird, C. M., \& Zepf, S. E. 1994, \aj, 108, 2348\\
Attridge, J. M. \& Herbst, W. \apjl, 398, L61 (AH)\\
Barnes, S. A., Sofia, S., Prosser, C. F., \& Stauffer, J. R. 1999, \apj, 516, 263\\
Bodenheimer, P. 1995, \araa, 33, 199\\
Bouvier, J., Bertout, C., Benz, W. \& Mayor, M. 1986, \aap, 165, 110\\
Bouvier, J. 1997, in ``Cool Stars in Clusters and Associations: Magnetic Activity and Age Indicators", Mem. della Societa' Astronomia Italiana, 68, 881  \\
Cameron, A.C., Campbell, C.G., \& Quaintrell, H. 1995 \aap, 298, 133\\
Choi, P. \& Herbst, W. 1996, \aj, 111, 283 (CH)\\
D'Antona, F. \& Mazzitelli, I. 1994, \apjs, 90, 467\\
Eaton, N. L., Herbst, W. \& Hillenbrand, L. A.  1995, \aj, 110, 1735\\
Edwards et al. 1993, \aj, 106, 372\\
Gebhardt, K. \& Beers, T. C. 1991, \aj, 383, 72\\
Ghosh, P. \& Lamb, F. K. 1979, \apj, 232, 259\\
Hartmann, L. W., Hewett, R., Stahler, S. \& Mathieu, R. D. 1986, \apj, 309, 275\\
Herbig, G. H. \& Terndrup, D. M. 1986, \apj, 307, 609\\
Herbst, W. \& Wittenmyer, R. 1996, \baas, 189, 4908\\
Hillenbrand, L. A. 1997, \aj, 113, 1733\\
Hillenbrand, L. A. \& Hartmann, L. W. 1998, \apj, 492, 540\\
Hillenbrand, L. A., Strom, S. E., Calvet, N., Merrill, K. M., Gatley, I., Makidon, R. B., Meyer, M. R., \& Skrutskie, M. F. 1998, \aj, 116, 1816\\
Horne, J. H. \& Baliunas, S. L. 1986, \apj,   302, 757\\
Jones, B.F. \& Walker, M. F. 1988, \aj, 95, 1755 (JW)\\
Johns-Krull, C. M., Valenti, J. A. \& Koresko, C. 1999, \apj, 516, 900\\
K\"ognil, A. 1991, \apjl, 370, L39\\ 
Krishnamurthi, A., Pinsonneault, M. H., Barnes, S. and Sofia, S. 1997, \apj, 480, 303\\
Li, J. \& Cameron, A. C. 1993 \mnras, 261, 766\\
Mahdavi, A. \& Kenyon, S.J. 1998, \apj, 497, 342\\
Mandel, G. N. \& Herbst, W. 1991, \apjl, 383, L75\\
Ostriker, E. \& Shu, F. 1995 \apj, 447, 813\\
Palla, F. \& Stahler, S. 1999 \apj, in press\\
Press, W. H., Flannery, B. P., Teukolsky, S. A., \& Vetterling, W. T. 1986, ``Numerical Recipes: The Art of Scientific Computing" (Cambridge U. Press, Cambridge), p. 489\\
Scargle, J. D. 1982, \apj,  263, 835\\
Sion, E. M. 1999, \pasp, 111, 532\\
Stahler, S. 1999, in ``Unsolved Problems in Stellar Evolution", ed. M. Livio, (Cambridge U. Press), in press\\
Stassun, K. G., Mathieu, R. D., Mazeh, T., \& Vrba, F. J. 1999, \aj, 117,294 (SMMV)\\
Stauffer, J. R. Caillault, J.-P., Gagne, M., Prosser, C.P., \& Hartmann, L.W. 1994, \apjs, 91, 625\\
Vogel, S. N. \& Kuhi, L. V. 1981, \apj, 245, 960\\
Vrba, F. J., Herbst, W. \& Booth, J. F. 1988, \aj, 96, 1032\\
Warren, W. H. \& Hesser, J. E. 1977, \apjs, 36, 497\\

\clearpage
\begin{figure}
\plotone{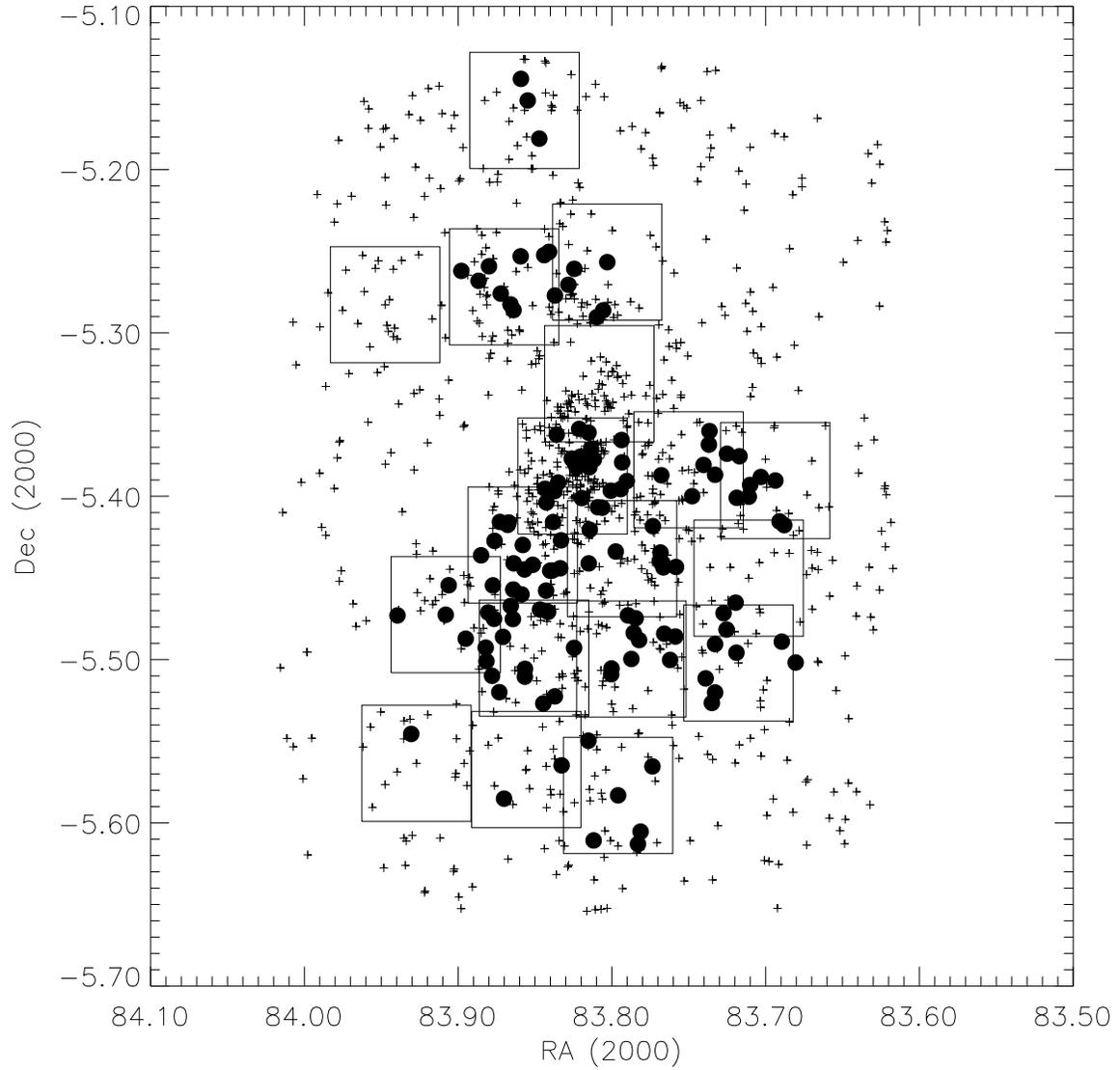}
\figcaption{The ONC. Stars in the Jones and Walker (1988) catalogue
are plotted as plus signs. Stars for which we have determined rotation
periods are shown as solid circles. The small rectangles outline the
individual fields surveyed at VVO, as detailed in Table 1.}
\end{figure}
 
\begin{figure} 
\plotone{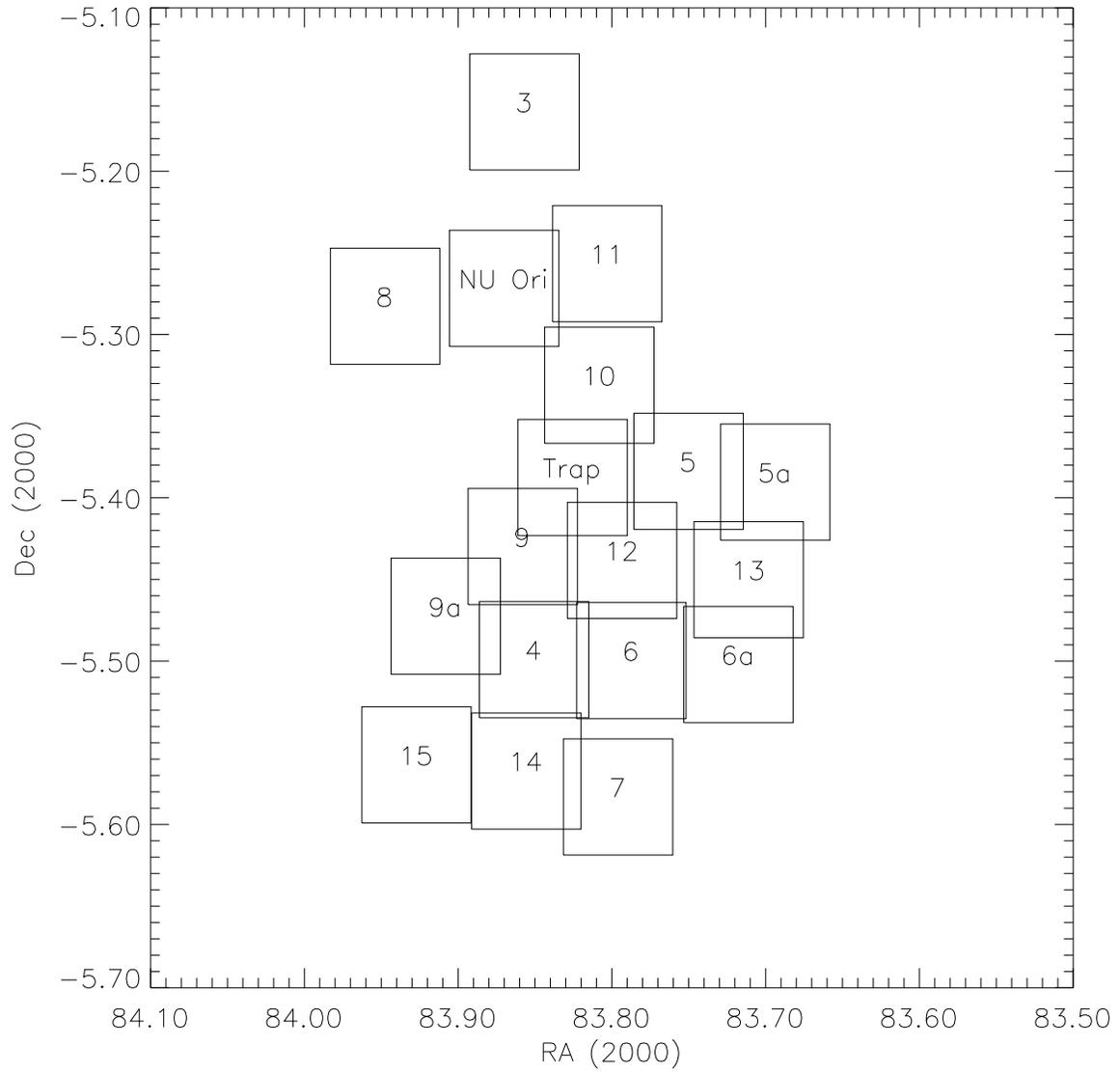}
\figcaption{The naming scheme for fields observed at VVO. See Table 1
for information on the epochs at which each field was monitored.}
\end{figure}  
 
\begin{figure} 
\epsscale{0.8}
\plotone{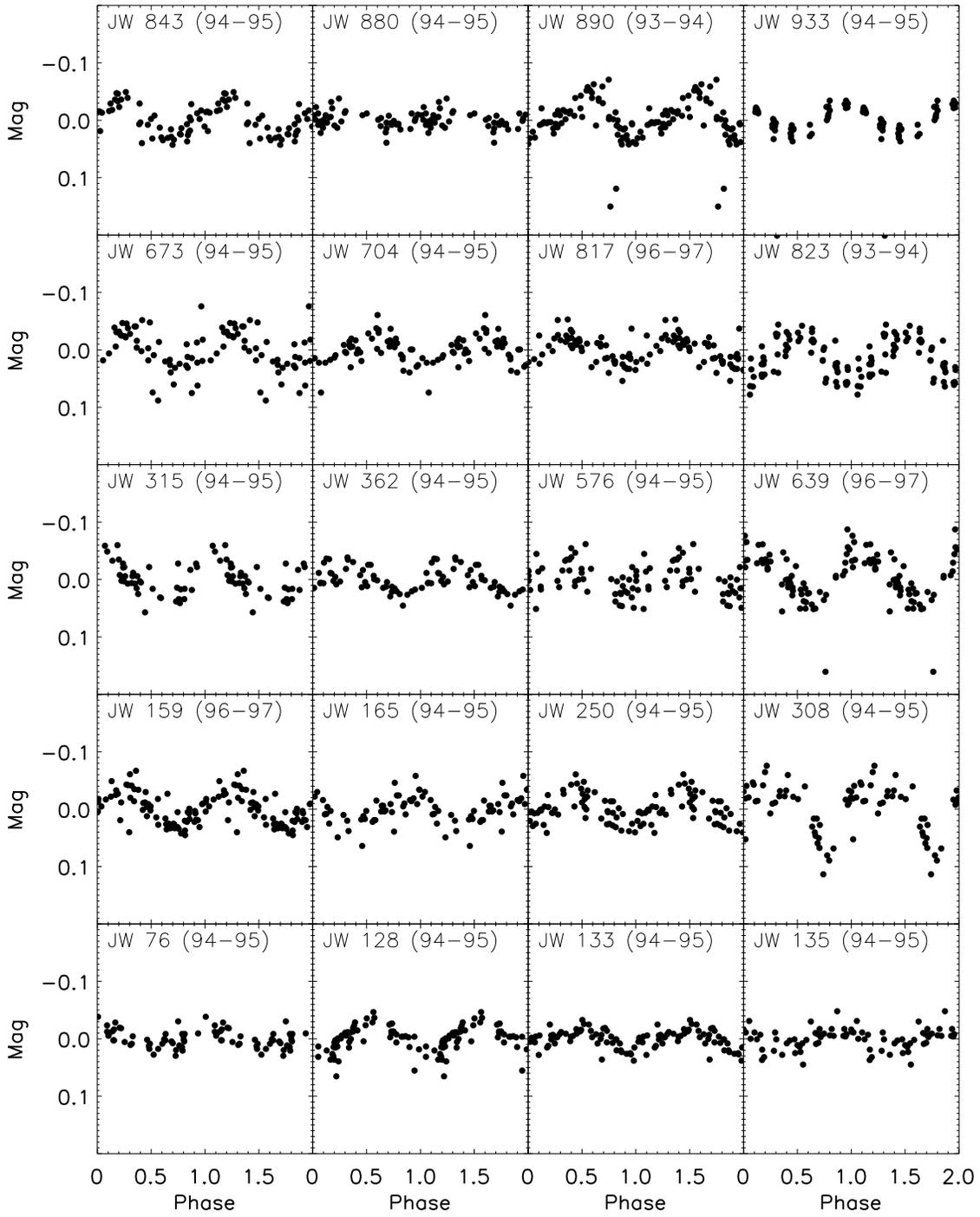}
\vskip 2cm
\figcaption{Light curves for stars in Table 2 which have not
previously been displayed by CH or AH.}
\epsscale{1.0}
\end{figure}  

\clearpage
\begin{figure} 
\figurenum{3}
\epsscale{0.8}
\plotone{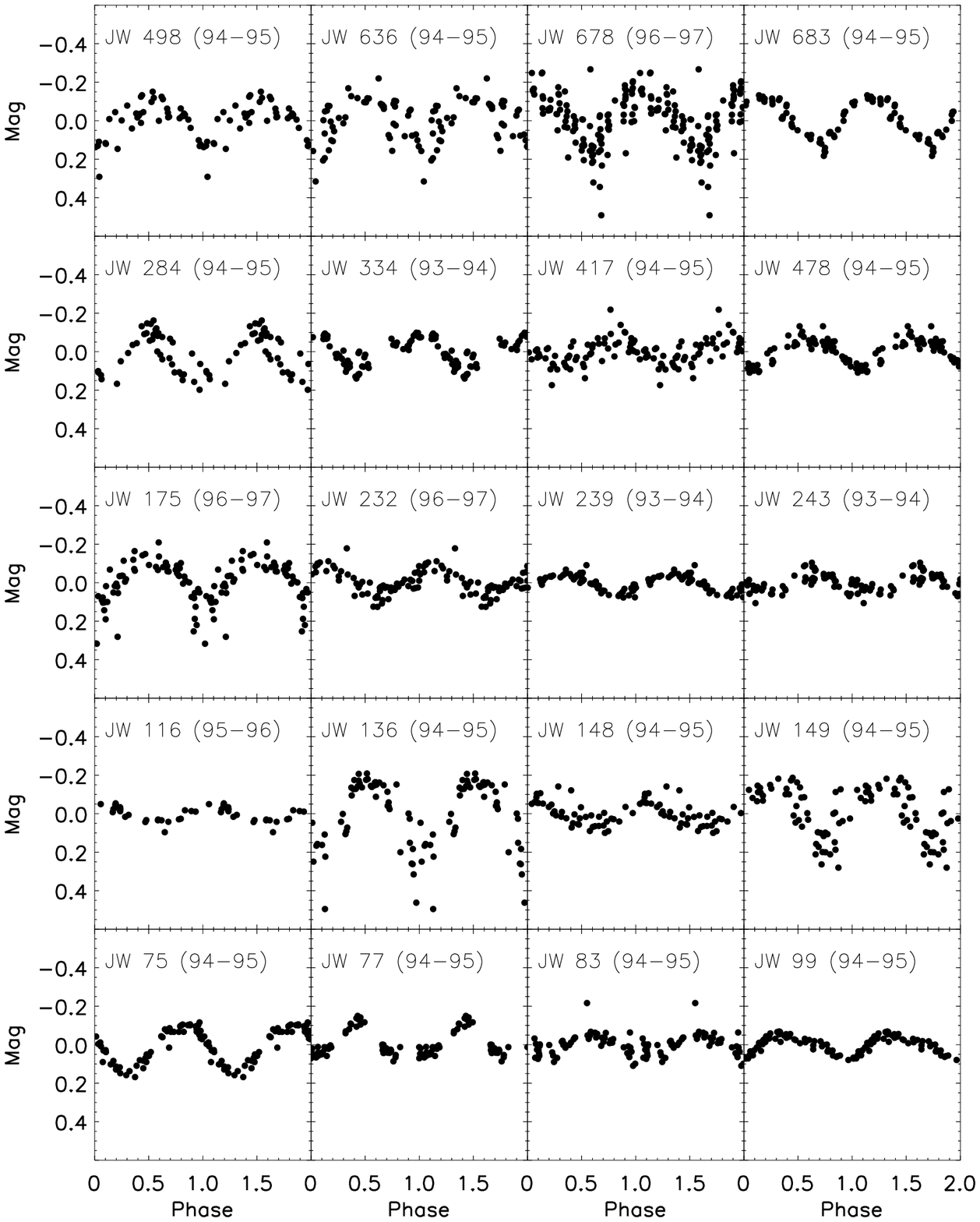}
\vskip 2cm
\figcaption{Light curves, continued.}
\epsscale{1.0}
\end{figure}  

\clearpage
\begin{figure} 
\figurenum{3}
\epsscale{0.8}
\plotone{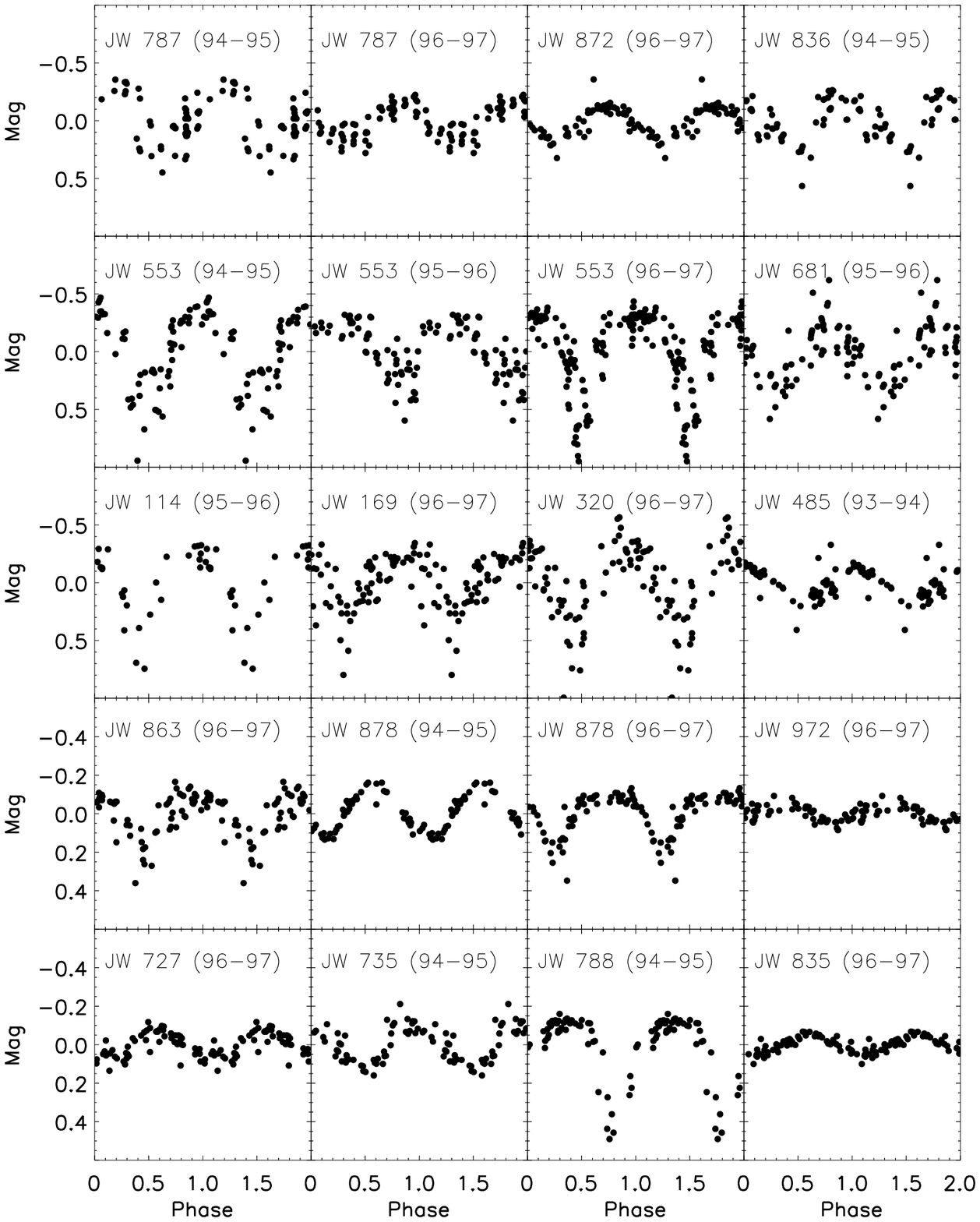}
\vskip 2cm
\figcaption{Light curves, continued.}
\epsscale{1.0}
\end{figure}  

\begin{figure} 
\plotone{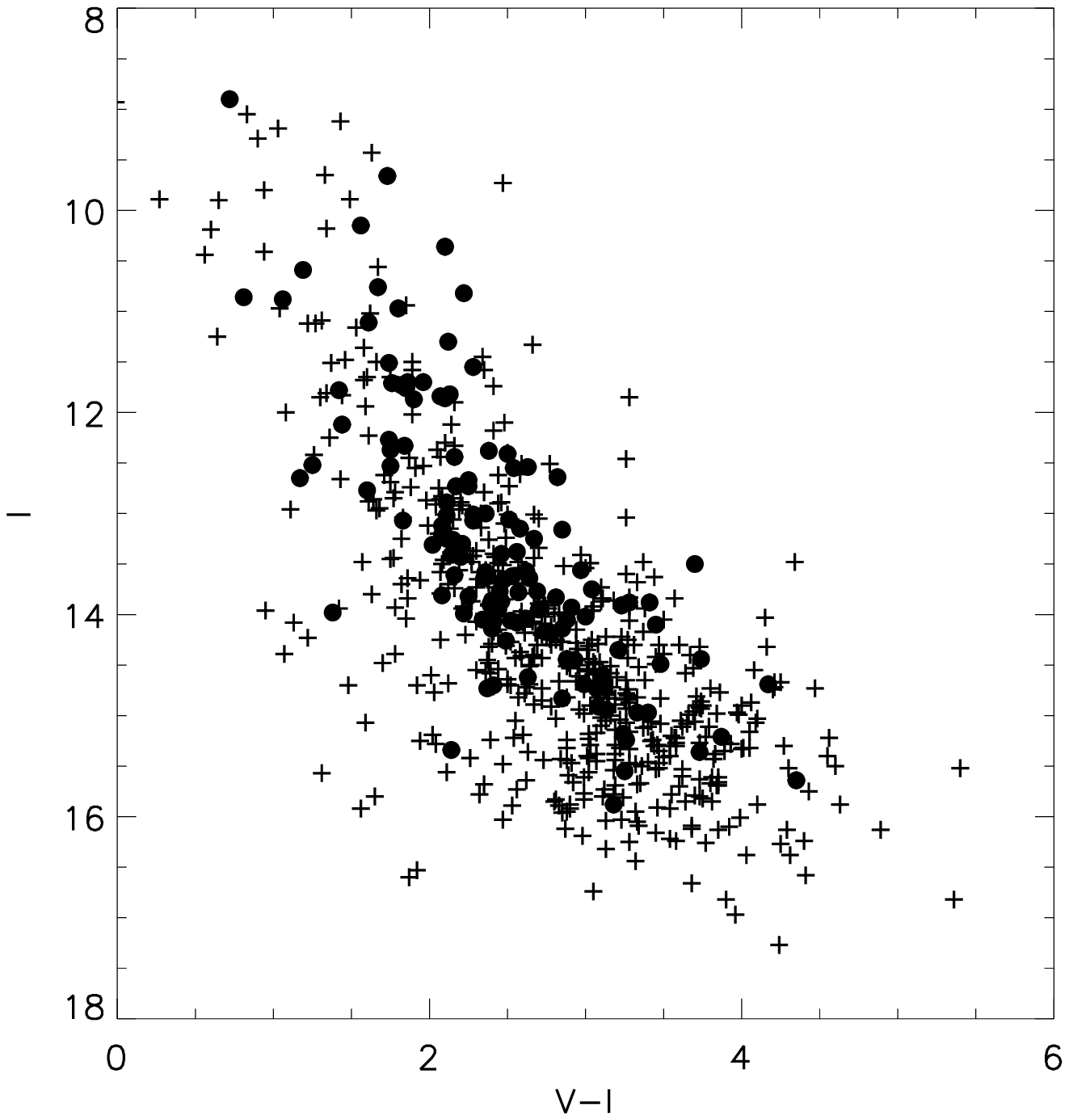}
\figcaption{Color-Magnitude diagram for JW stars monitored by
us. Solid circles indicate periodic stars and plus signs indicate
those not found to be periodic.}  
\end{figure}

\begin{figure} 
\epsscale{0.6} 
\plotone{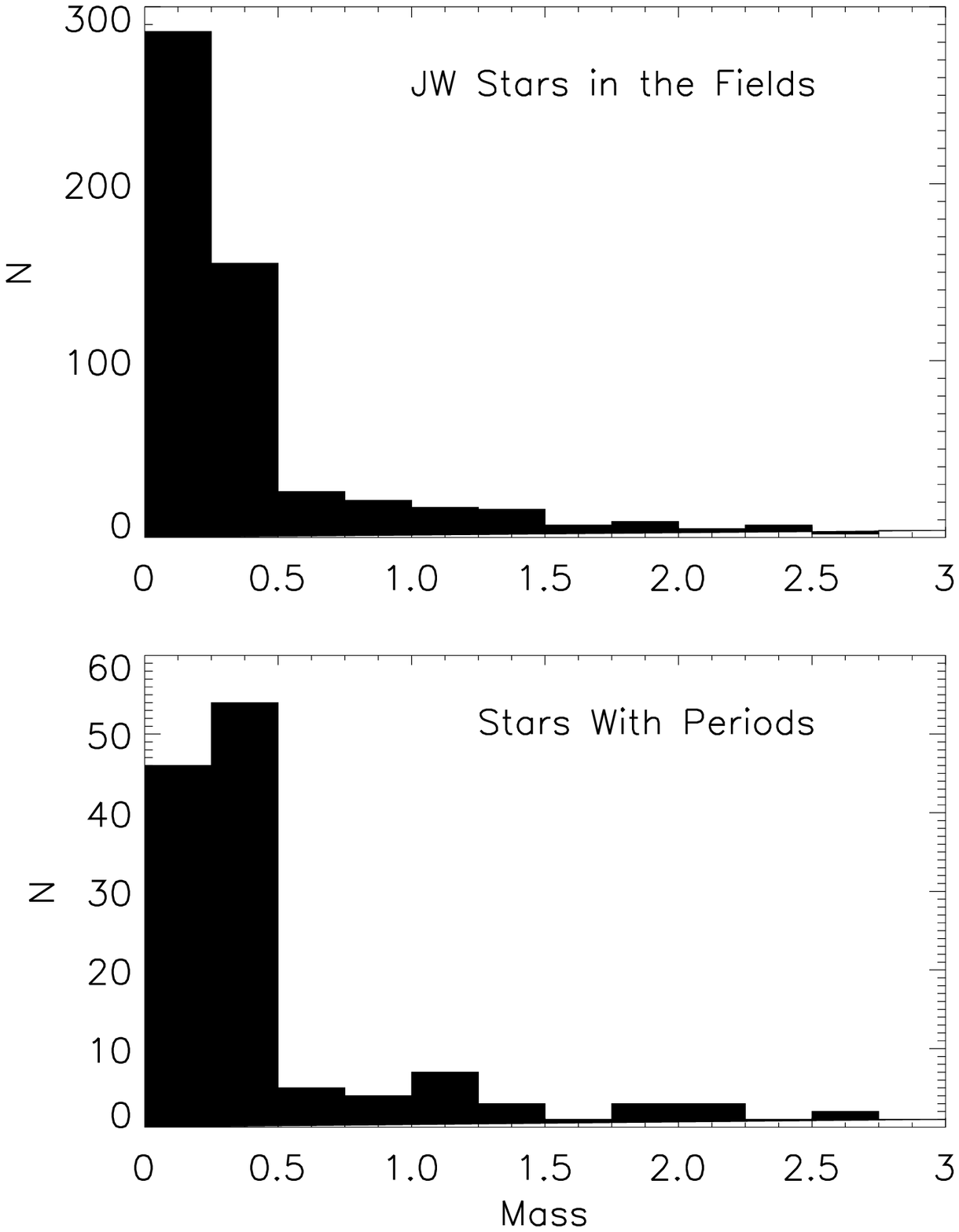}
\figcaption{The mass distribution for all JW stars in our fields (top
panel) and for the stars for which rotation periods have been found
(bottom panel).}
\epsscale{1.0}
\end{figure}  

\begin{figure} 
\plotone{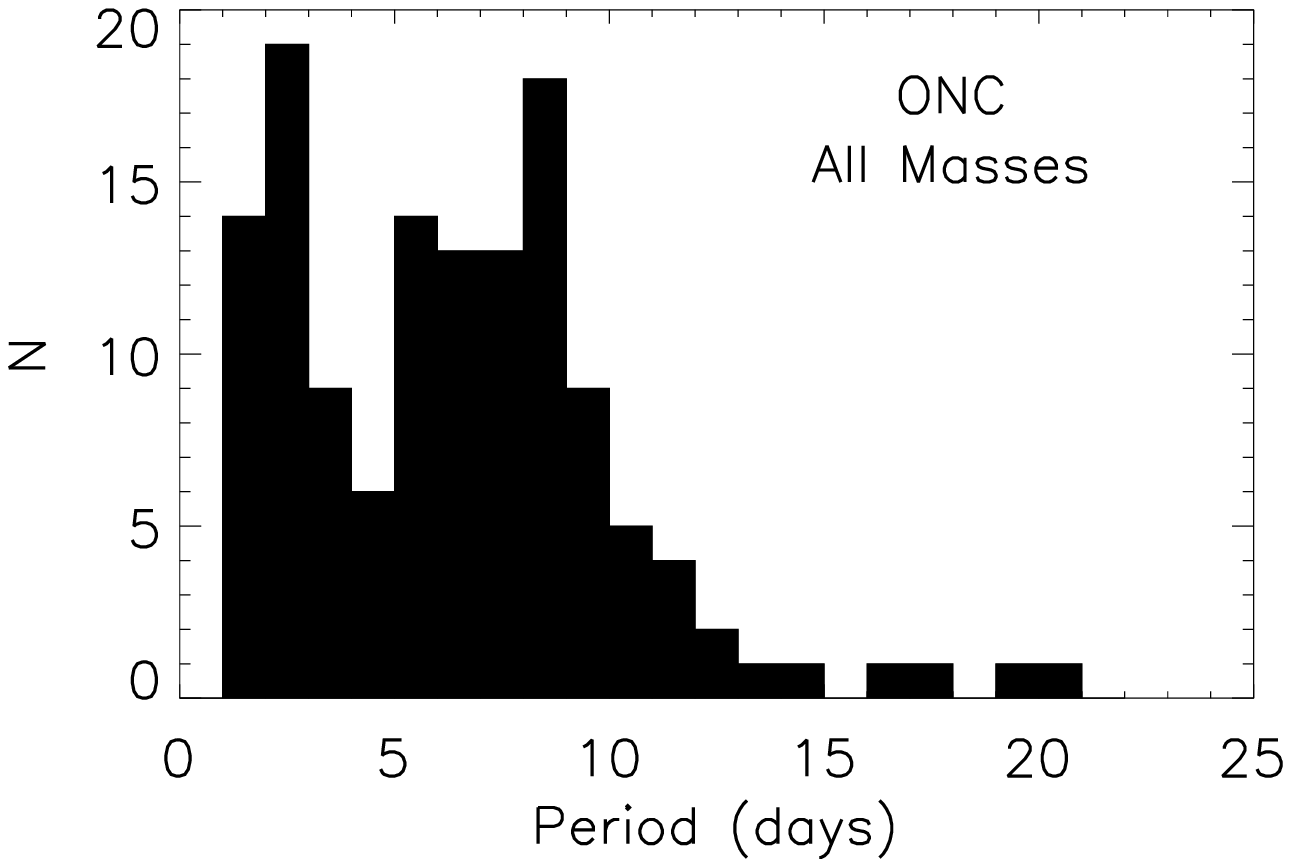}
\figcaption{The frequency distribution of rotation periods for stars
of all mass in the ONC. One star, with a period of 34 days, lies
outside the boundaries of the figure.}
\end{figure}  

\begin{figure}
\epsscale{0.6} 
\plotone{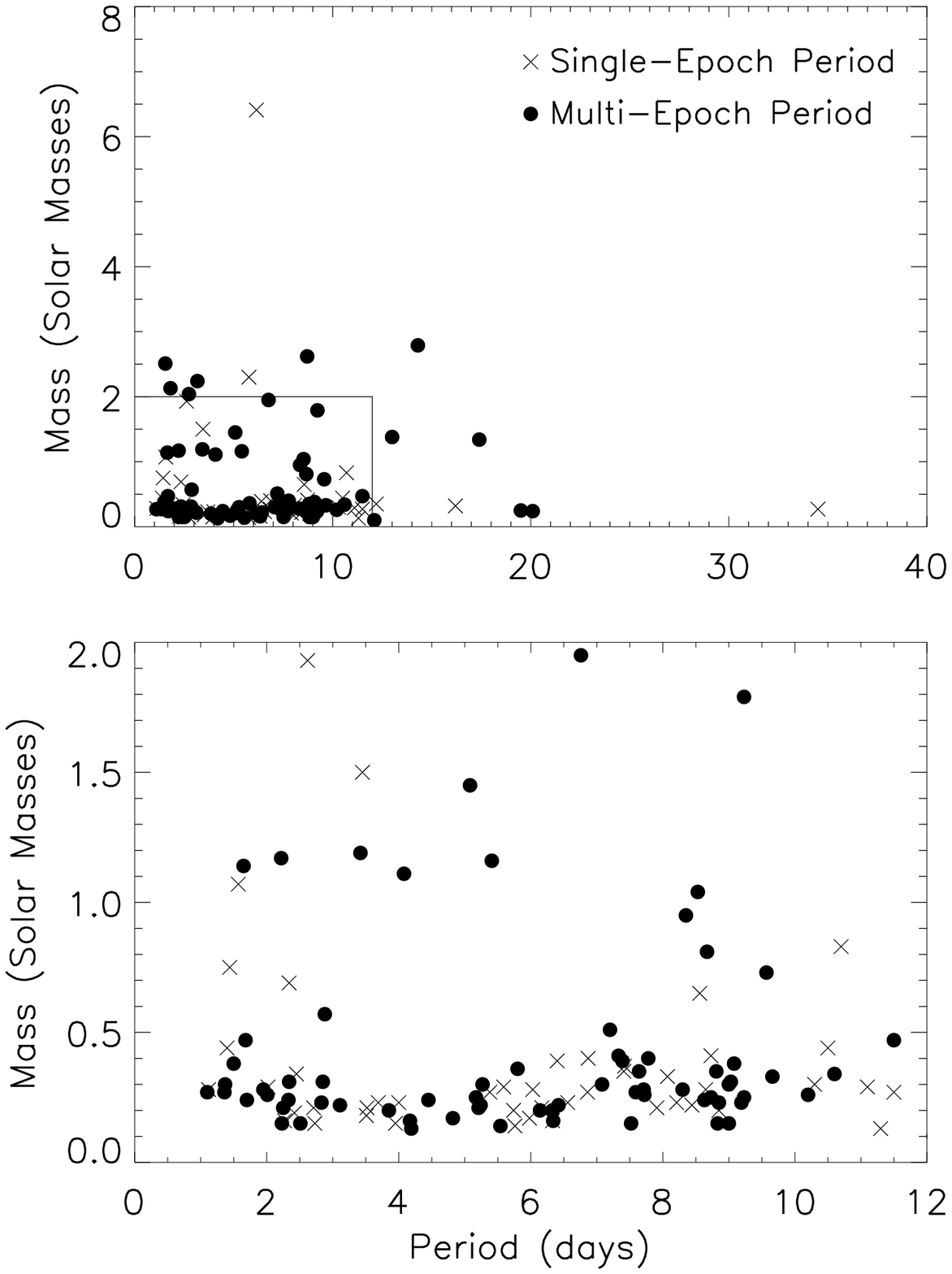}
\vskip 2cm
\figcaption{Mass versus rotation period. Solid circles are
multi-epoch periods; crosses are single-epoch periods. The top panel
shows the entire data set and the bottom panel is an expanded view of
the boxed region.}
\epsscale{1.0}
\end{figure}  

\begin{figure}
\epsscale{0.6} 
\plotone{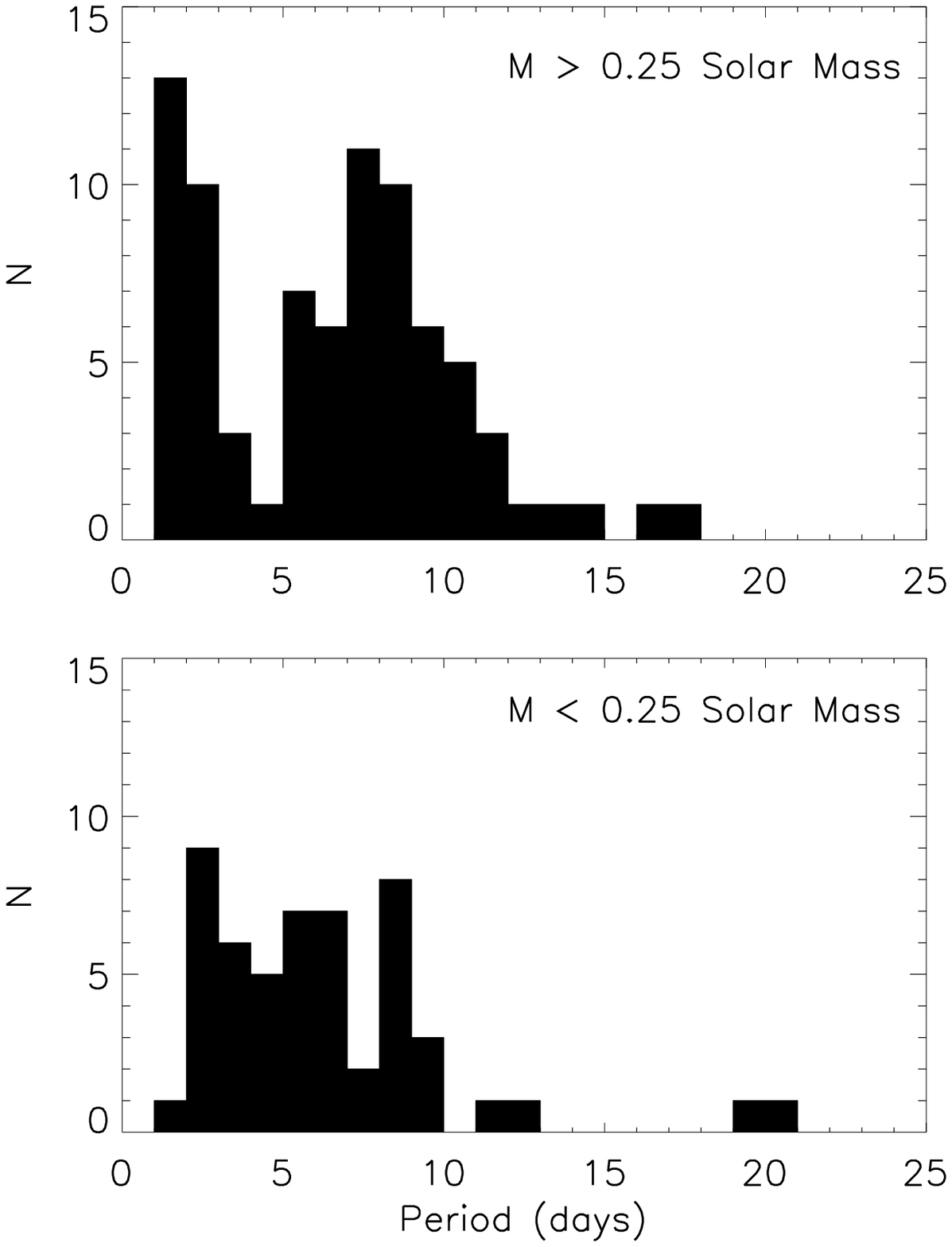}
\figcaption{Rotation period distributions for more massive stars (top
panel) and less massive stars (lower panel).}
\epsscale{1.0}
\end{figure}  

\begin{figure} 
\plotone{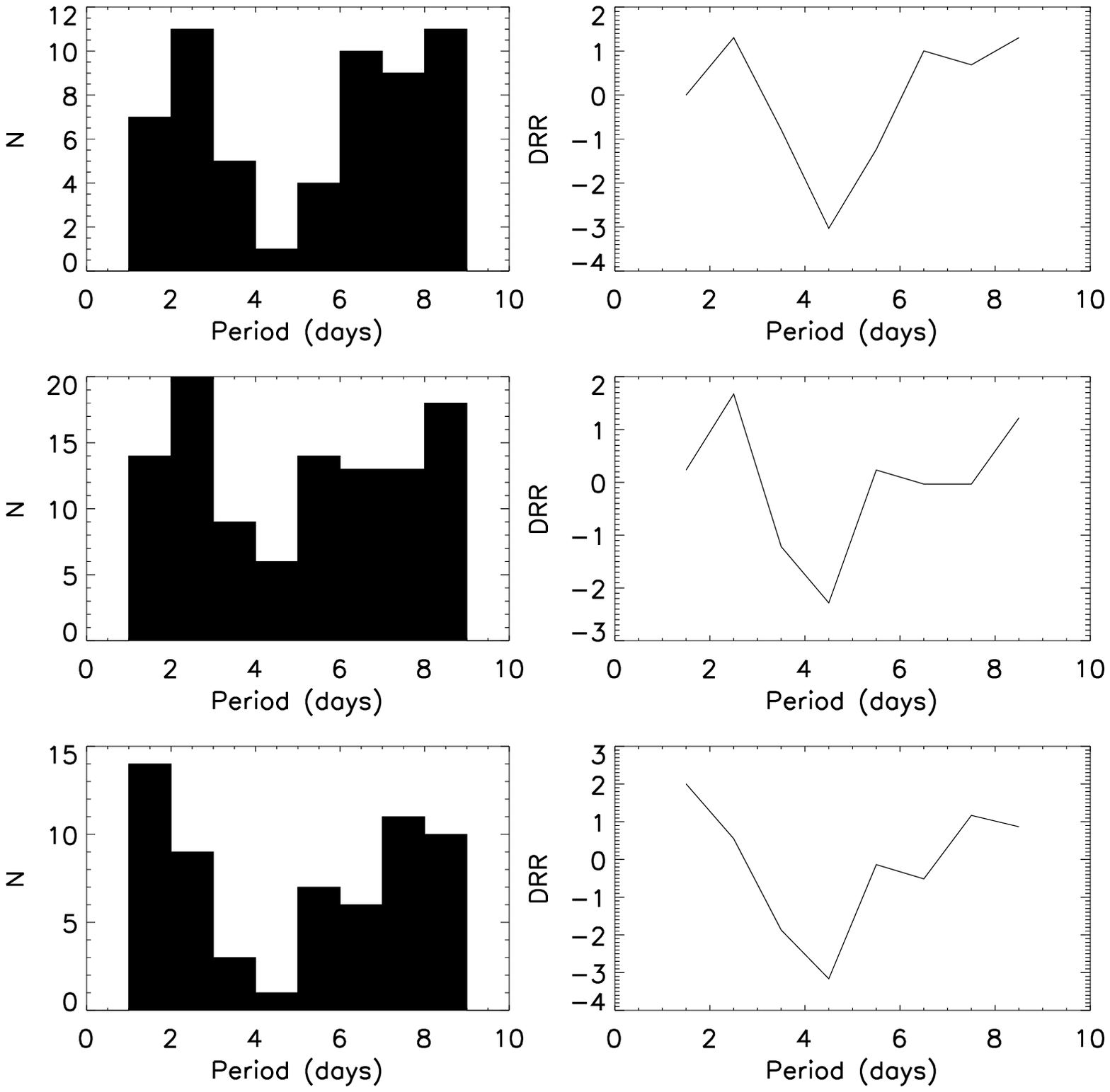}
\figcaption{Period distributions (right) and DRR test results (left)
for three samples: CH (top), all masses (middle), stars with M $>$
0.25 M$_{\odot}$ (bottom).}
\end{figure}  

\begin{figure} 
\plotone{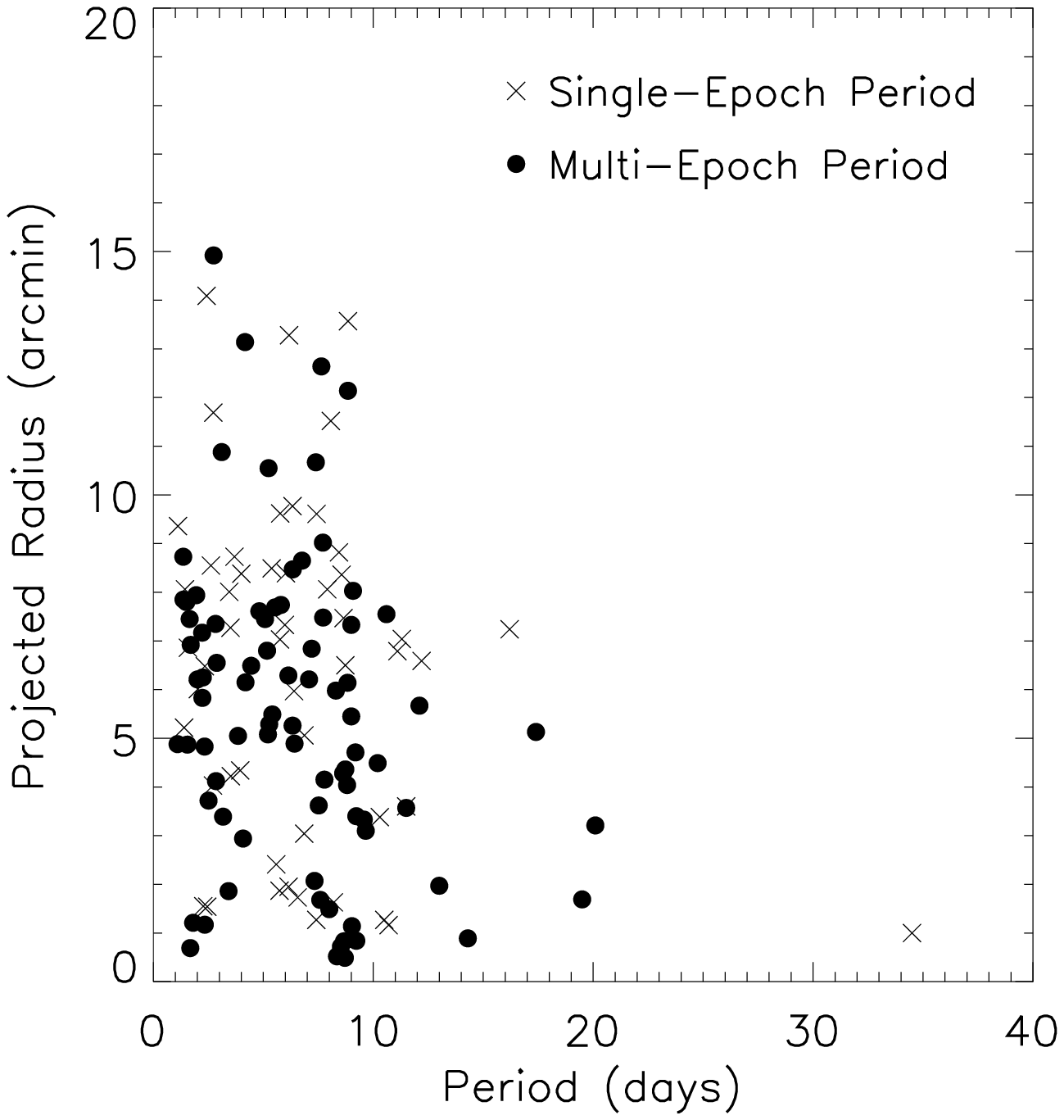}
\figcaption{ Projected Radius (in arc-minutes) from the center of the
ONC versus Rotation Period. Solid circles are multi-epoch periods and
crosses are single-epoch periods.}
\end{figure}  

\begin{figure} 
\plotone{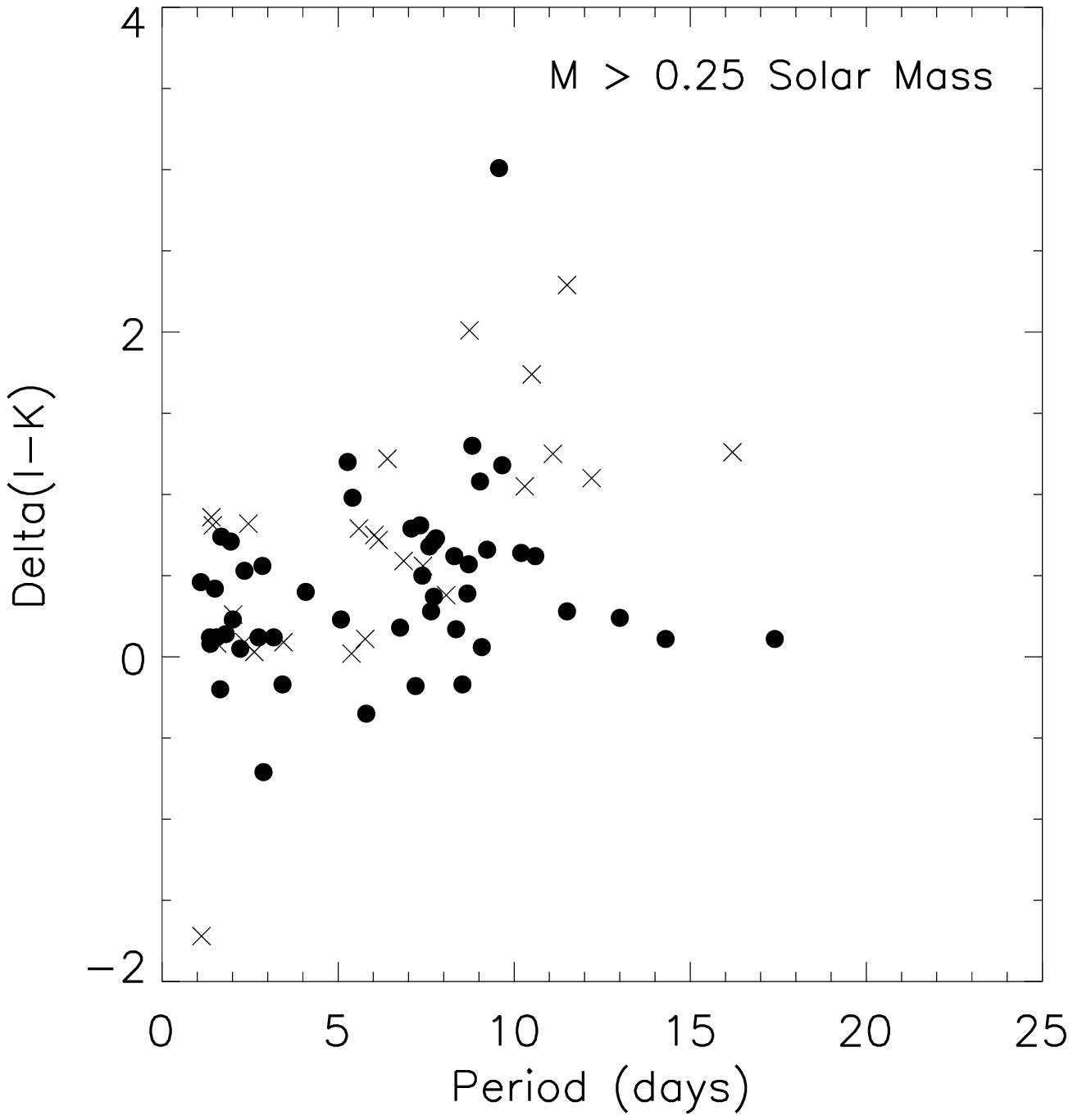}
\figcaption{Excess I-K emission from Hillenbrand (1997) versus
rotation period for stars with M $>$ 0.25 M$_{\odot}$. Solid circles
are multi-epoch periods and crosses are single-epoch periods.}
\end{figure}  

\begin{figure} 
\plotone{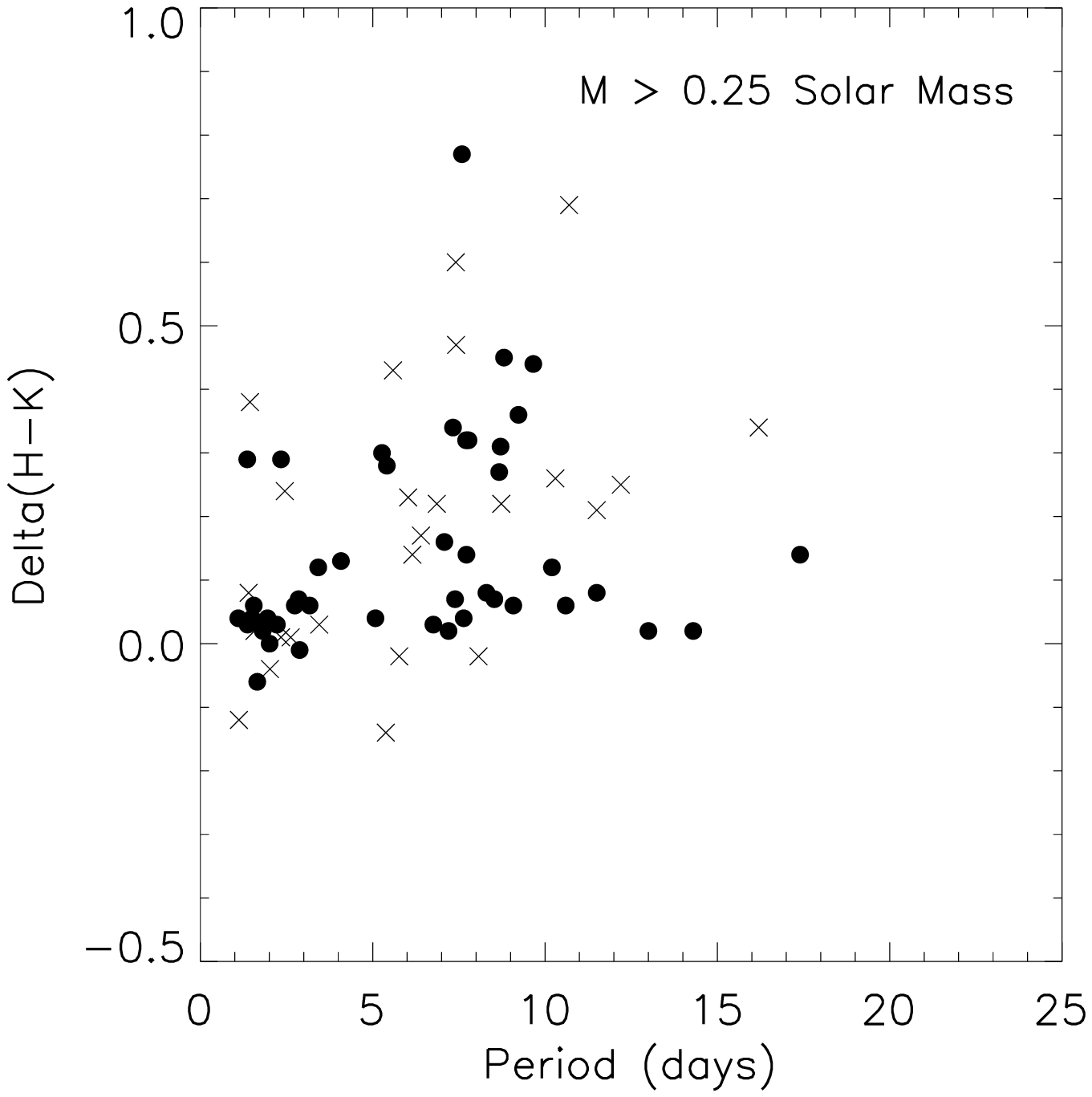}
\figcaption{Excess H-K emission from Hillenbrand (1999) versus
rotation period for stars with M $>$ 0.25 M$_{\odot}$. Solid circles
are multi-epoch periods and crosses are single-epoch periods.}
\end{figure}  

\begin{figure} 
\plotone{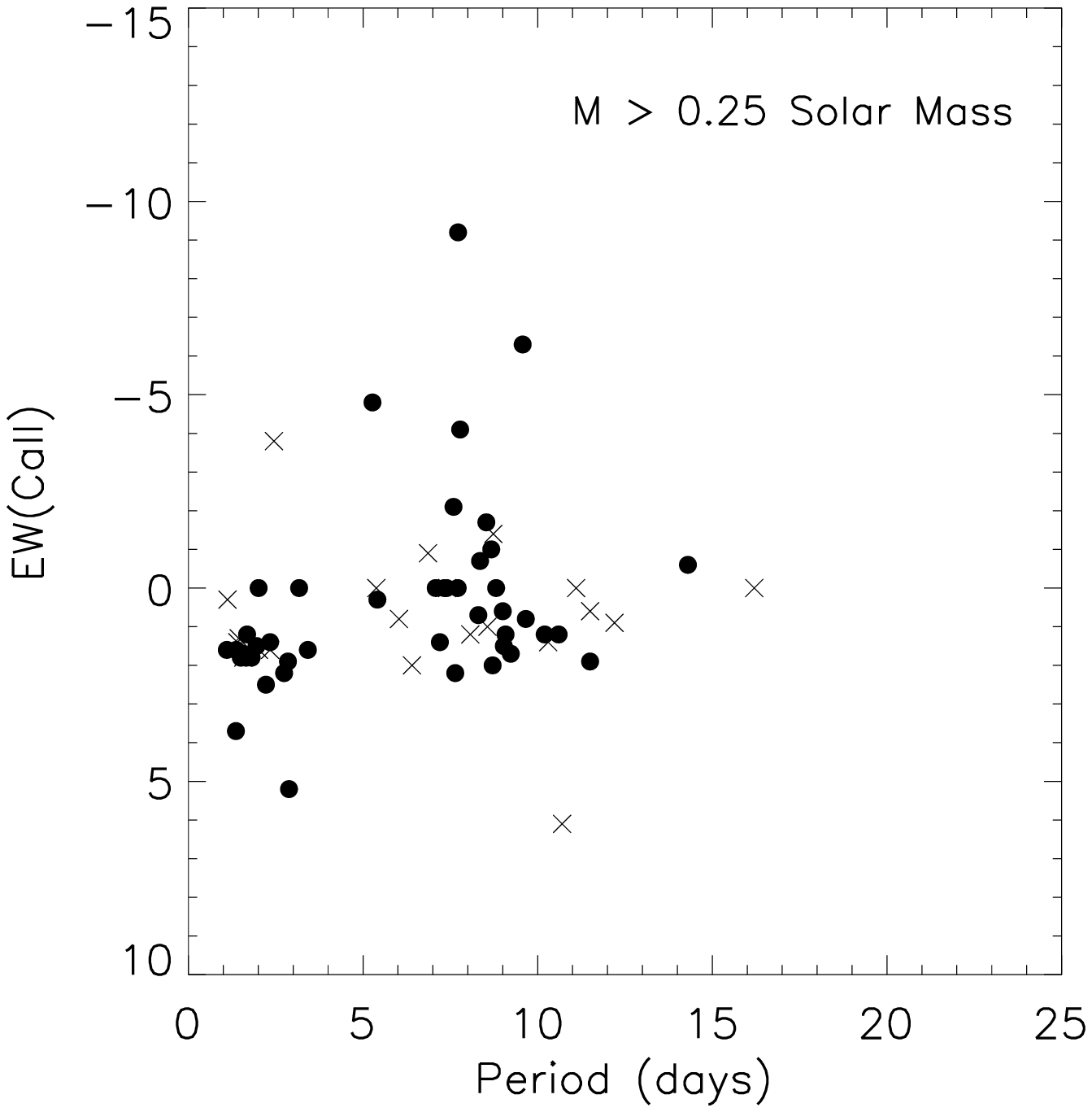}
\figcaption{Equivalent width of Calcium II emission from Hillenbrand
(1997) versus rotation period for stars with M $>$ 0.25
M$_{\odot}$. Solid circles are multi-epoch periods and crosses are
single-epoch periods.}
\end{figure}  

\begin{figure} 
\plotone{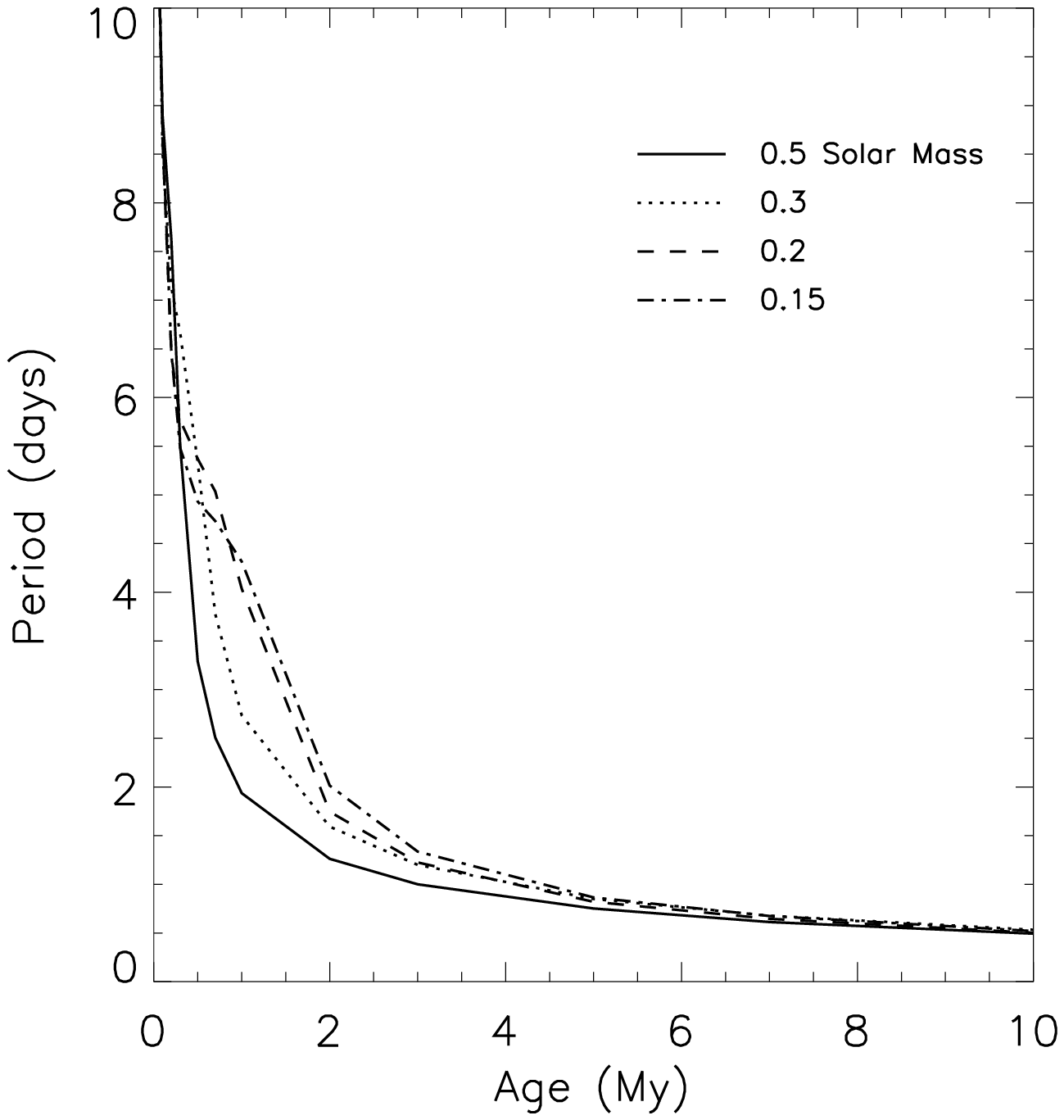}
\figcaption{The variation of rotation period with time for stars of
different masses, based on the models of DM94. The starting point is a
period of 10 days at an age of 0.07 My.}
\end{figure}  

\begin{figure} 
\plotone{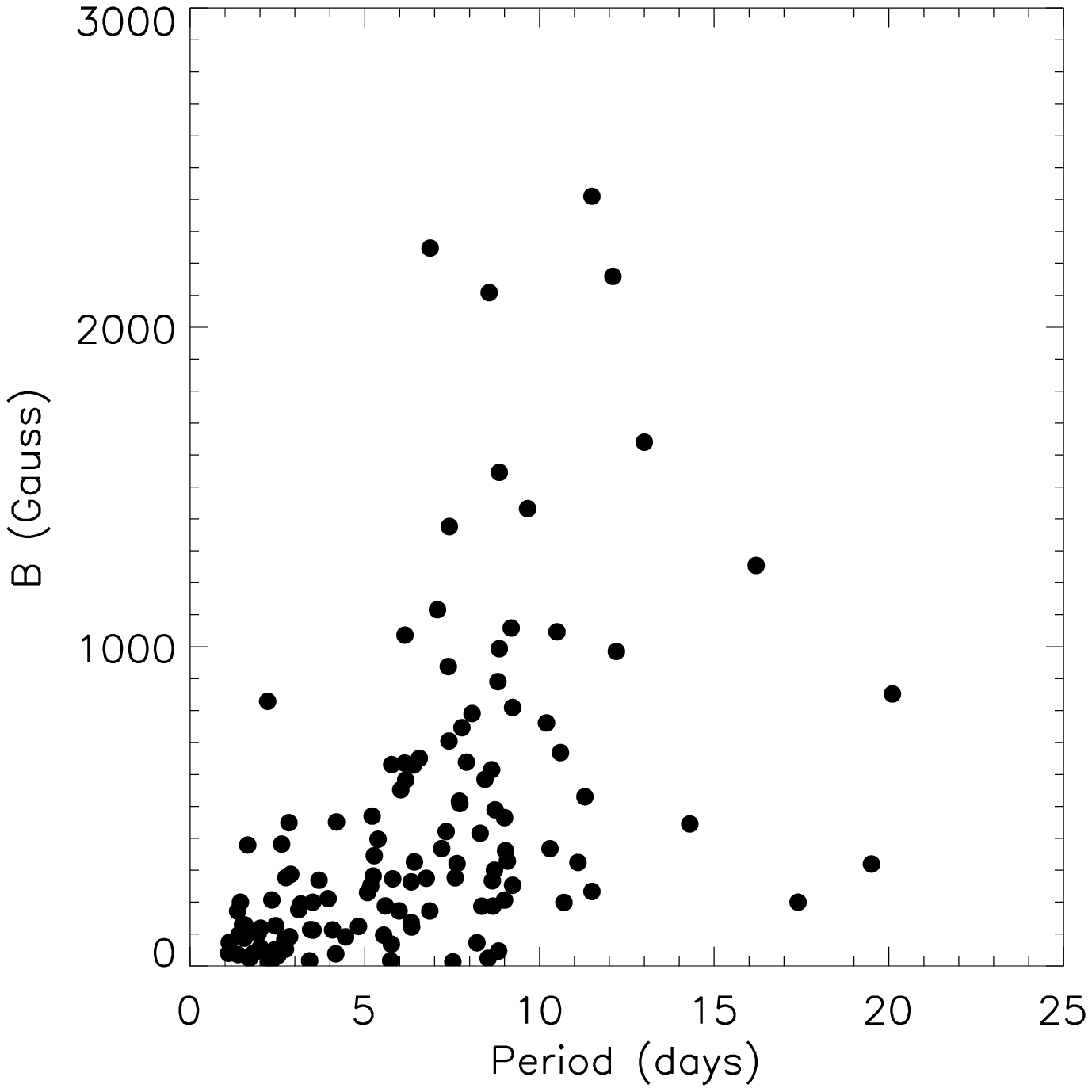}
\figcaption{The mean equatorial magnetic field strength, B (in Gauss),
predicted by the disk-locking theory of Ostriker and Shu (1995)
assuming a mass accretion rate of 10$^{-8} \dot M_{\odot} yr^{-1}$
versus rotation period.}
\end{figure}  

\begin{figure} 
\plotone{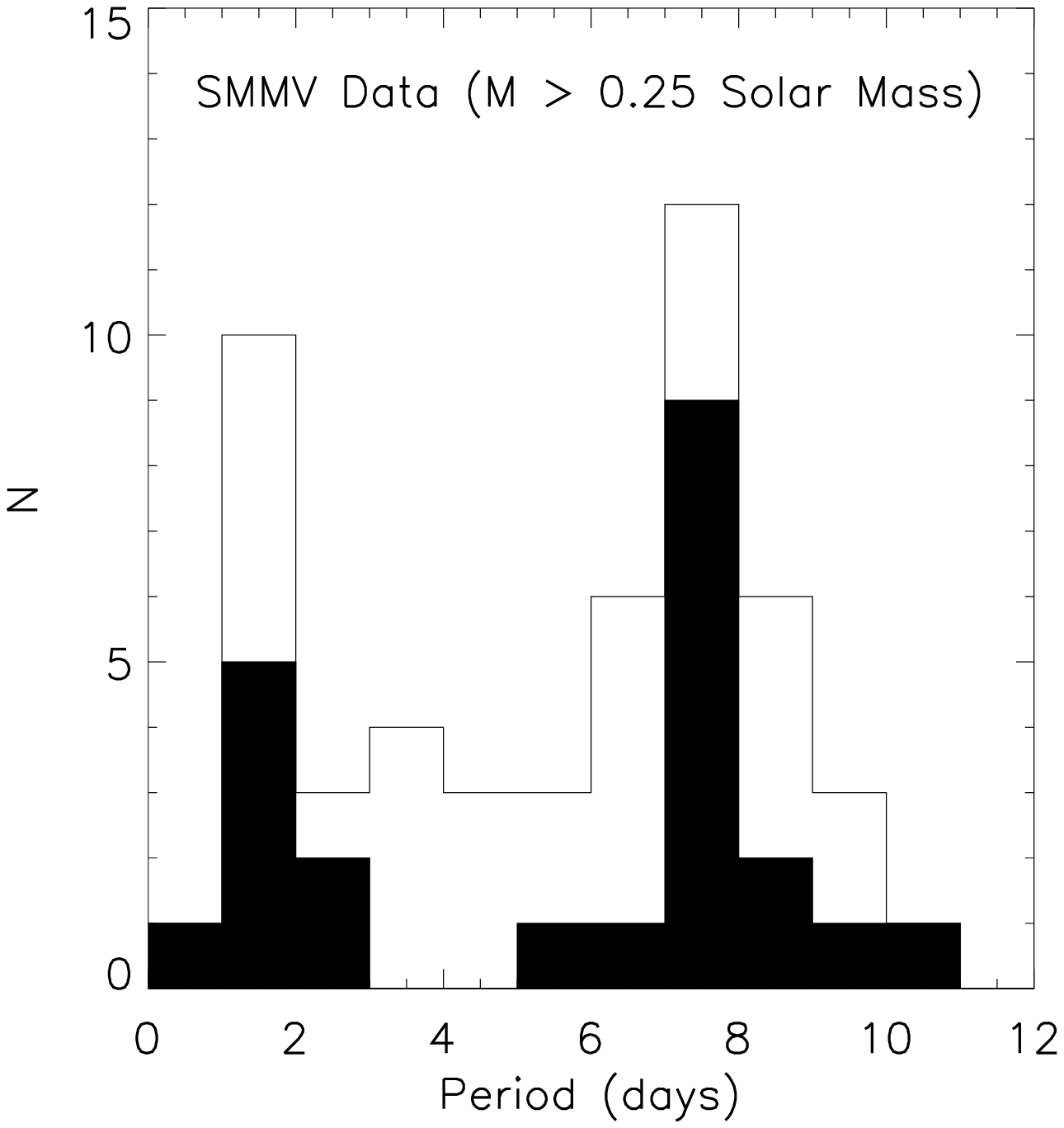}
\figcaption{The frequency distribution of stars with M $>$ 0.25
M$_{\odot}$ and rotation periods determined by SMMV. The solid portion
of the histogram is for stars whose rotation periods are confirmed in
this study.}
\end{figure}  

\begin{figure} 
\plotone{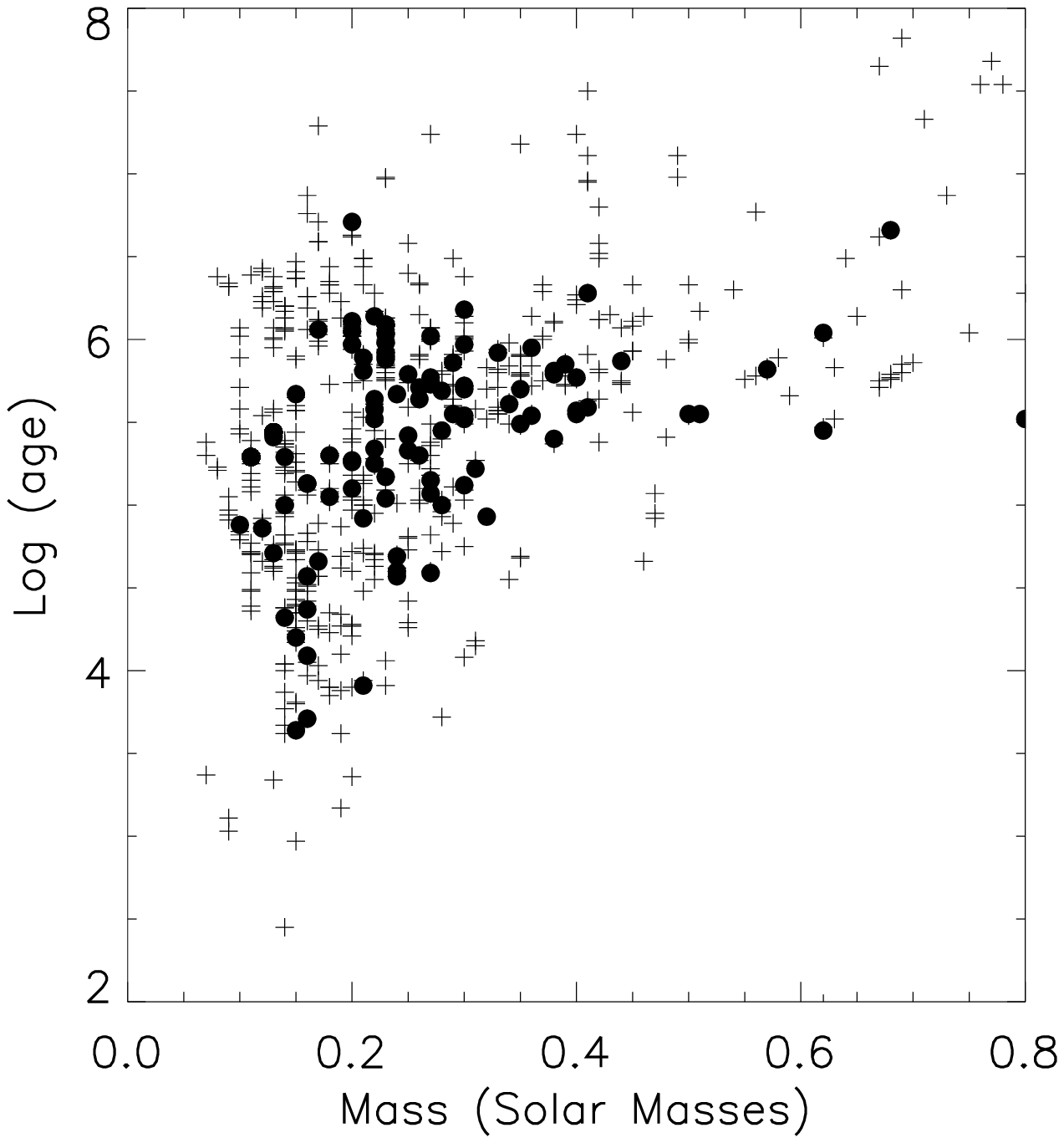}
\figcaption{Age, as determined by Hillenbrand (1997) based on the
models of DM94, versus mass. Plus signs are JW stars without known
rotation periods. Solid circles are stars with periods from SMMV. This
figure illustrates the correlation that exists between ``age" and mass
in the samples studied by SMMV.}
\end{figure}  

\end{document}


\footnotesize
\begin{deluxetable}{cccccccccccc}
\tablenum{2}
\tablewidth{0pt}
\tablecaption {Periods For Stars in the ONC}
\tablehead{\colhead{JW}&\colhead{90-91}&\colhead{91-92}&\colhead{92-93}&\colhead{93-94}&\colhead{94-95}&\colhead{95-96}&\colhead{96-97}&\colhead{97-98}&\colhead{98-99}&\colhead{Adopt}&\colhead{SMMV}}
\startdata
65&	&	&	7.44&	&	7.34&	&	&	&	&	7.39&7.04\nl
75&	&	&	&	&	3.45&	&	&	&	&	3.45\nl
76&	&	&	&	&	6.33&	&	&	&	&	6.33\nl
77&	&	&	&	&	1.50&	1.50&	&	&	&	1.50&1.50\nl
83&	&	&	&	&	7.72&	&	&	&	&	7.72&7.40\nl
99&	&	&	&	&	1.70&	1.70&	&	&	&	1.70&1.69\nl
114&	&	&	&	&	&	8.73&	&	&	&	8.73\nl
116&	&	&	&	&	&	2.34&	&	&	&	2.34\nl
128&	&	&	&	&	8.83&	&	&	&	&	8.83&8.35\nl
133&	&	&	&	&	2.02&	&	&	&	&	2.02\nl
135&	&	&	&	&	3.69&	&	&	&	&	3.69\nl
136-137&&	&	&	&	8.65&	&	&	&	&	8.65\nl
145&	&	12.15&	&	&       12.02&	&	&    	&	&	12.1\nl
148&	&	&	&	&	1.37&	&	&	&	&	1.37&1.37\nl
149&	&	&	&	&	2.83&	&	&	&	&	2.83&2.76\nl
157&	&	17.25&	&	17.4&    17.6&  &  	&	&	&	17.4\nl
158&	&	&	&	&	&	1.95&	1.94&	&	&	1.95&1.92\nl
159&	&	&	&	&	&	&	1.12&	&	&	1.12\nl
165&	&	&	&	&	5.77&	&	&	&	&	5.77\nl
167&	&	&	1.40&	&	&	&	&	&	&	1.40\nl
169&	&	&	&	&	3.86&	&	3.83&	&	&	3.85&3.82\nl
174&	&	&	1.36&	&	1.36&	&	&	&	&	1.36&1.36\nl
175&	&	&	&	9.16&	9.24&	9.18&	9.16&	&	&	9.19\nl
191\tablenotemark{a}&	&	&	8.54&	&	&	&	&	8.67&	8.69&	8.63&8.82\nl
220&	&	&	2.32&	2.33&	&	&	&	&	2.32&	2.33\nl
222&	&	&	5.17&	5.16&	&	&	&	&	&	5.17&5.06\nl
232&	&	&	&	&	&	&	5.09&	&	5.07&	5.08\nl
239&	&	4.46&	&	4.43&	&	&	&	&	&	4.45&4.44\nl
243&	&	&	&	10.2&	&	&	&	&	&	10.2&10.1\nl
245&	&	&	8.63&	&	8.83&	&	&	&	&	8.73&9.46\nl
248&	&	6.86&	&	&	&	&	&	&	&	6.86\nl
250&	&	&	&	&	2.71&	&	&	&	&	2.71\nl
254&	&	&	3.52&	&	&	&	&	&	&	3.52\nl
275&	&	&	20.4&	19.8&	20.3&	20.2&	19.8&	&	&	20.1\nl
284&	&	&	&	&	3.11&	&	&	&	&	3.11&3.09\nl
308&	&	&	&	&	4.17&	&	&	&	&	4.17&4.20\nl
311&	&	6.14&	6.13&	6.13&	&	&	6.15&	&	&	6.14&6.13\nl
315&	&	&	&	&	8.85&	&	&	&	&	8.85\nl
320&	&	&	&	&	5.43&	&	5.39&	&	&	5.41\nl
326&	&	&	6.40&	&	&	&	&	&	&	6.40\nl
330&	 &       1.57&	&	&	&	&	&	&	&	1.57\nl
334&	&	&	&	5.27&	&	&	&	&	&	5.27&5.32\nl
337&	&	&	19.3&	&	&	&	&	19.7&	&	19.5\nl
345\tablenotemark{b}&	&	&	&	8.21&	&	&	&	&	&	8.21\nl
347&	7.31&	7.32&	7.33&	7.38&	7.33&	&	&	7.39&	&	7.33&7.33\nl
352&	7.98&	7.99&	8.03&	7.95&	7.95&	8.11&	7.97&	&	&	8.00\nl
362&	&	&	&	&	2.73&	&	&	&	&	2.73\nl
365&	&	&	4.08&	&	&	&	&	&	4.09&	4.08\nl
378&	9.05&	9.01&	&	&	&	&	&	&	&	9.03\nl
379\tablenotemark{c}&	&	5.65&	&	&	&	&	5.59&	&	&	11.3&11.3\nl
381&	&	&	16.2&	&	&	&	&	&	&	16.2\nl
388&	&	9.08&	9.08&	&	&	&	&	&	9.10&	9.08&7.78\nl
406&	&	&	2.25&	&	&	&	&	&	&	2.25&2.26\nl
417&	&	&	&	&	7.41&	&	&	&	&	7.41\nl
437&	&	&	2.33&	2.34&	2.34&	&	&	&	&	2.34\nl
439&	&	&	8.30&	&	&	&	&	&	&	8.30&8.35\nl
454&	&	8.69&	8.70&	8.62&	&	&	&	&	8.66&	8.67\nl
466&	&	6.17&	&	&	&	&	&	&	&	6.17\nl
470&	&	&	10.69&	&	&	&	&	&	&	10.7\nl
478&	&	&	&	&	5.74&	&	&	&	&	5.74\nl
479&	&	8.74&	&	&	&	&	8.68&	&	&	8.71\nl
481\tablenotemark{d}&	&	&	6.56&	&	&	&	&	&	&	6.56\nl
485&	&	&	&	9.63&	9.68&	&	&	&	&	9.66&9.46\nl
498&	&	&	&	&	7.42&	&	&	&	&	7.42\nl
526&	&	1.68&	1.69&	1.68&	1.68&	&	&	&	1.68&	1.68\nl
536&	&	&	9.21&	9.24&	&	&	&	&	&	9.23\nl
544&	&	3.43&	3.43&	3.42&	3.41&	3.42&	&	&	3.40&	3.42\nl
553&	&	&	&	&	8.39&	8.34&	8.33&	&	&	8.35\nl
559&	&	&	&	&	&	&	&	2.27&	&	2.27\nl
563&	&	&	5.80&	&	&	&	&	&	&	5.80&6.12\nl
567&	8.58&	8.53&	&	8.48&	&	&	8.54&	&	8.52&	8.53\nl
576\tablenotemark{e}&	&	&	&	&	2.01&	&	1.08&	&	&	2.01\nl
589&	14.32&	14.32&	14.28&	14.28&	14.38&	&	14.36&	&	&	14.3\nl
601&	&	2.22&	&	&	&	&	&	&	2.23&	2.22\nl
636&	&	&	&	&	5.59&	&	&	&	&	5.59\nl
639&	&	&	&	&	&	&	5.26&	&	5.22&	5.24&5.16\nl
641&	&	&	3.16&	3.18&	&	&	&	&	3.16&	3.17\nl
648\tablenotemark{f}&	&	&	&	34.51&	&	&	&	&&34.5\nl
660&	6.15&	&	&	&	&	&	&	&	&	6.15\nl
664&	&	7.20&	7.19&	&	&	&	&	&	7.23&	7.20&7.63\nl
669&	&	&	&	1.81&	&	&	1.81&	&	1.81&	1.81\nl
673&	&	&	&	&	1.44&	&	&	&	&	1.44\nl
678\tablenotemark{g}&	&	&	&	&	&	&	&	13.02&	12.93&	13.0\nl
681&	&	&	&	&	&	10.46&	&	&	&	10.5\nl
683&	&	&	&	11.5&	11.5&	&	&	&	&	11.5\nl
691&	&	6.34&	&	&	&	&	&	&	&	6.34&6.29\nl
695&	&	11.53&	&	&	&	&	&	&	&	11.5\nl
704&	&	&	&	&	6.87&	&	&	&	&	6.87\nl
710&	7.53&	&	7.63&	7.60&	7.56&	7.57&	7.65&	&	&	7.59&7.66\nl
716&	&	&	3.95&	&	&	&	&	&	&	3.95\nl
717&	&	&	4.00&	&	&	&	&	&	&	4.00\nl
721\tablenotemark{h}&	&	&	&	2.45&	&	&	&	&	&	2.45\nl
727&	&	&	&	&	&	&	6.03&	&	&	6.03\nl
731&	&	7.63&	7.64&	&	&	&	&	&	&	7.64&7.70\nl
735&	&	&	&	&	5.21&	&	&	&	&	5.21&5.21\nl
758&	&	2.51&	2.51&	2.51&	&	&	&	&	&	2.51&2.49\nl
765&	&	&	2.42&	&	&	&	&	&	&	2.42\nl
784-787&&	&	&	&	8.97&	&	9.04&	&	&	9.00&7.74\nl
786&	&	8.79&	8.80&	8.84&	&	&	&	&	8.80&	8.81\nl
788\tablenotemark{i}&	&	&	&	&	4.82&	&	&	&	&	4.82&4.76\nl
790&	&	2.74&	2.74&	& 	&	&	&	&	&	2.74\nl
792&	&	10.33&	&	&	&	&	&	&	&	10.3\nl
794&	&	&	2.62&	&	&	&	&	&	&	2.62\nl
795&	&	1.56&	1.56&	&	1.54&	&	&	&	&	1.55\nl
811&	&	&	11.05&	&	&	&	&	&	&	11.1\nl
813&	&	2.85&	2.85&	2.84&	2.84&	&	&	&	&	2.85&2.84\nl
815&	&	6.45&	6.40&	6.40&	&	&	&	&	6.52&	6.42&6.43\nl
817\tablenotemark{j}&	&	&	&	&	4.14&	&	2.23&	&	2.24&	2.23\nl
819&	&	&	5.76&	&	&	&	&	&	&	5.76\nl
823&	&	9.00&	&        9.00&	&	&	&	&	&	9.00\nl
826&	&	&	9.58&	9.59&	9.55&	&	&	&	&	9.57\nl
830&	&	9.26&	9.27&	9.16&	&	&	&	&	&	9.23\nl
835&	&	&	&	&	&	&	8.85&	&	&	8.85&8.64\nl
836&	&	&	&	&	12.18&	&	&	&	&	12.2\nl
837&	&	10.32&	10.77&	&	&	&	&	&	10.78&	10.6&8.69\nl
839&	&	7.55&	7.51&	7.54&	7.49&	&	&	&	&	7.52\nl
843\tablenotemark{k}&	&	&	&	&	5.38&	&	&	&	&	5.38&0.84\nl
850&	&	7.78&	7.78&	&	&	&	&	&	&	7.78\nl
855&	&	7.02&	&	&	7.25&	&	6.97&	&	&	7.08&7.00\nl
860&	&	6.31&	6.35&	6.33&	&	&	&	&	6.36&	6.33&6.30\nl
863&	&	&	&	&	&	&	7.91&	&	&	7.91\nl
866&	&	6.77&	6.75&	&	&	&	&	&	&	6.76\nl
872&	&	&	&	&	4.19&	&	4.20&	&	&	4.19\nl
878&	&	&	&	&	5.52&	&	5.55&	&	5.55&	5.54&5.51\nl
880&	&	&	&	&	3.51&	&	&	&	&	3.51\nl
890\tablenotemark{l}&	&	&	&	1.03&	&	&	&	&	1.16&	1.10\nl
892&	&	8.56&	&	&	&	&	&	&	&	8.56\nl
907&	&	&	1.65&	1.65&	1.65&	&	&	&	&	1.65&1.65\nl
914&	&	7.71&	7.71&	&	&	&	&	&	&	7.71\nl
930&	&	&	2.87&	2.88&	2.88&	&	&	&	&	2.88&2.79\nl
933&	&	&	&	&	5.98&	&	&	&	&	5.98\nl
972&	&	&	&	&	&	&	8.07&	&	&	8.07\nl
984\tablenotemark{m}&	&	&	8.46&	8.41&	&	&	&	&	&	8.44&1.13\nl
H3128\tablenotemark{n}&	&	&	&	2.47& 	&	&	&	&	&	2.47\nl
\enddata

\tablenotetext{a}{JW 191. A period of 4.33 days with 0.03 $<$ FAP $<$ 0.01 was found in 91-92. This is
one-half of the likely true period; presumably the star had spots
in two hemispheres during this epoch, similar to what has been observed for V410 Tau (Vrba, Herbst \& Booth 1987).}

\tablenotetext{b}{JW 345. There is a large peak in the combined power spectrum (from all seasons) of this star near 21 days.}

 \tablenotetext{c}{JW 379. See the text for a discussion of the difference between our result and SMMV.}

 \tablenotetext{d}{JW 481. The period  possibility near 1.18 days has slightly larger power in the combined power spectrum and could be the true period rather than an alias.}

 \tablenotetext{e}{JW 576. Both periods are very close to integral days. We adopted 2.01 days 
based on the quality of the light curve. Because this is so close to the one day
sampling interval, it is particularly uncertain and requires confirmation.}

\tablenotetext{f}{JW 648. Large peaks in the combined power spectra are seen at 3.36 days and 1.42 days and a smaller peak at 34.5 days.}

\tablenotetext{g}{JW 678. This star was first found by combining the power spectra obtained
at all epochs, which yielded a large peak at the reported period. It also has
an FAP $<$ 0.001 in the 95-96 epoch. It clearly has periodic variation at
about that period at three other epochs: 92-93, 93-94 and 94-95.}

\tablenotetext{h}{JW 721. A period of 11.37 days with 0.03 $<$ FAP $<$ 0.01 was found in 91-92.}

\tablenotetext{i}{JW 788. This star has an unusual light curve shape reminiscent of a W UMa star.}

\tablenotetext{j}{JW 817. The period of 4.14 days is inconsistent with v sin i for this star, whereas the
2.23 day period is about what one would predict based on its likely radius and measured
v sin i, assuming that sin i is close to 1.}

\tablenotetext{k}{JW 843. A period of 10.88 days with 0.03 $<$ FAP $<$ 0.01 was found in 91-92. 
Note the similarity to twice the period found in 94-95. See the text for a discussion of the difference between our result and SMMV.}

\tablenotetext{l}{JW 890. A period that close to 1 day would be hard to detect, given our
sampling interval. Nonetheless, it may be real, but requires confirmation.}

\tablenotetext{m}{JW 984. See the text for a discussion of the difference between our result and SMMV.}

\tablenotetext{n}{H 3128. This star does not have a JW identification. It is star 3128 in Hillenbrand's (1997) compilation.}

\end{deluxetable}